\RequirePackage{fix-cm}
\documentclass{svjour3}                     
\smartqed  
\usepackage{graphicx}
\usepackage{amsmath,amssymb}
\usepackage{latexsym}
\usepackage{ntheorem}
\usepackage{ifthen}
\usepackage{enumerate}
\usepackage{color}
\usepackage{hyperref}
\usepackage{newcent}
\usepackage{tikz,tkz-tab}
\usepackage{mathtools}
\usepackage{dsfont}
\newcommand\scalemath[2]{\scalebox{#1}{\mbox{\ensuremath{\displaystyle #2}}}}
\newtheorem{assumption}[theorem]{Assumption}
\newcommand{\essup}[1][\Pc]{\mathrm{ess\,sup}_{#1}}
\newcommand{\essinf}[1][\Pc]{\mathrm{ess\,inf}_{#1}}
\newcommand{\supp}[1][\Pc]{\mathrm{ supp\,}_{#1}}

\def \one{\hbox{I\hskip-.60em 1}}

\def\1{\mbox{\bf 1}}

\def \R{\mathbb{R}}
\def \Q{\mathbb{Q}}
\def \QQ{\mathbb{Q}}
\def \N{\mathbb{N}}
\def\Ac{{\cal A}}
\def\Bc{{\cal B}}
\def\Cc{{\cal C}}

\def\Pc{{\cal P}}
\def\Oc{{\cal O}}

\def\Qc{{\cal Q}}
\def\Rc{{\cal R}}
\def\Sc{{\cal S}}
\def\Xc{{\cal X}}
\def\Kc{{\cal K}}

\def \a{\alpha}

\def \l{\lambda}

\def \p{\pi}

\def \f{\varphi}
\def \o{\omega}
\def \O{\Omega}

\def \e{\varepsilon}
\begin{document}

\title{Quasi-sure essential supremum and applications to finance
}
\subtitle{}


\author{Laurence Carassus 
}


\institute{L. Carassus \at Laboratoire de Math\'{e}matiques de Reims, UMR9008 CNRS et Universit\'{e} de Reims Champagne-Ardenne, France 
              \email{ laurence.carassus@univ-reims.fr}           
           }

\date{Received: date / Accepted: date}

\maketitle

\begin{abstract}
When uncertainty is modelled by a set of non-dominated and non-compact probability measures, a notion of essential supremum for a family of real-valued functions is developed in terms of upper semi-analytic functions. 
We show how the properties postulated on the initial functions  carry over to  their  quasi-sure essential supremum. 
We  propose various  applications to financial problems with frictions. We analyse super-replication and prove a  bi-dual characterization of the super-hedging cost. We also study a weak no-arbitrage condition called Absence of Instantaneous Profit (AIP) under which prices are finite.  This requires new results on the aggregation of quasi-sure statements.

\end{abstract}

\section{Introduction}
We place ourselves in the so-called quasi-sure setting where the unique probability measure prevailing on the state space is replaced by a set of measures $\Pc$  which is neither supposed to be dominated, nor supposed to be compact. One may wonder 
whether it is still possible to define an essential supremum with respect to $\Pc$ for a collection of real-valued functions $X_i$ where  $i$ belongs to an arbitrary index set $I$.
 The answer is yes if we choose the right notion of measurability  for the $X_i$. Then, our essential supremum inherits the same type of measurabilty.

Our quasi-sure essential supremum is well-suited for financial applications. Indeed, the reliance on a single probability measure has long been questioned in the economic literature  and is often referred to as Knightian uncertainty, in reference to \cite{Kni}. In a financial context, it is  called model-risk and also has a long history, we refer to \cite{OW18} and the references  therein.  
Here, we consider  the  quasi-sure (or multiple-priors) approach developed in \cite{BN}. As the set of probability measures can vary between a singleton  and all the probability measures on a given space, this formulation encompasses  a wide range of  settings, including the classical one.  As the set of priors is not assumed to  be dominated, this has raised challenging  mathematical  questions  and has lead to the development of innovative  tools such as quasi-sure stochastic analysis, non-linear expectations and G-Brownian motions. On these topics,  we refer among  others to \cite{DM06}, \cite{pg07} or  \cite{SoToZa11}.  
In  \cite{Bart216}, the Choquet theorem is generalized to conditional sublinear expectation for upper semi-analytic functions.  

We first present the construction and the mathematical properties of the quasi-sure essential supremum. 
We show that an appropriate notion of upper semi-analytic essential supremum exists for a collection indexed by $I$ of real-valued functions $( \o,\widetilde{\o}) \mapsto X_i( \o,\widetilde{\o})$ whenever the union of $\supp[\Pc]X_i$ the quasi-sure support of  $X_i$ is assumed to be an upper semi-analytic measurable set-valued mapping.
This measurability condition on $\cup_{i \in I}\supp[\Pc]X_i$  is in particular true if $I$ is countable, the $X_i$ are upper semi-analytic and the graph of $\Pc$ is an analytic set. All this conditions will be true for our financial applications. 
We prove under suitable assumptions that the quasi-sure essential supremum of $h(X)$ is the supremum of $h$ on $\supp[\Pc]X,$ generalizing the result of  \cite{BCL} obtained when there is only one prior. We also show that some properties postulated on $\Phi( \o,\widetilde{\o}, \cdot): \theta \mapsto \Phi( \o,\widetilde{\o},\theta)$ carry over to $\f$ the  quasi-sure essential supremum  of $\Phi$ as a function of $\theta:$ 
$\theta \mapsto \f(\o,\theta).$ We prove subtle results on the measurability of 
 the epigraphical set-valued mapping  of $\Phi$ and also on the joint measurability of $\f.$ This will allow us  to do measurable selection. 

Then, we consider a general one period financial market with $d$ risky assets and one perfectly liquid num\'eraire, whose cost is set equal to one. The trading costs of the $d$ risky assets are given by real-valued cost functions  $ \Sigma=(\Sigma_0, \Sigma_1).$  This means that buying a portfolio $(\theta^0, \theta)\in  \R^{d+1}$ at time $0$ (resp. $1$)
and in the state $\o$  (resp. $(\o, \tilde \o)$) costs $\theta^0 + \Sigma_0(\o,  \theta)$ (resp. $\theta^0 + \Sigma_1 (\o, \tilde \o,\theta)$) units of cash. 
This is a very general formulation which contains frictionless markets, markets with transaction costs or bid-ask spreads  and also markets with illiquidity effects. 
We  also consider a contingent claim $Z$. We remark that the projection on $\R$ of the epigraph of $\Phi ^{\Sigma}+Z$, where 
$\Phi ^{\Sigma}(\o, \tilde \o,\theta)= \Sigma_0(\o,  \theta) +  \Sigma_1 (\o,\tilde \o, -\theta)$ is the cost of the strategy $\theta,$ 
is in fact the set of super-hedging prices  of the contingent claim $Z$. 
This allows to le the set of super-hedging prices as well as its infimum called  the  super-hedging cost. In particular we give conditions for this set to be closed, i.e. for the  super-hedging cost to be a super-hedging price. Then, we provide a dual characterization of the super-hedging prices and a bi-dual characterization of the super-hedging cost using a duality based on the cost functions  rather than martingale measures. 
In the case of a frictionless market, the duality is the one of Fenchel-Legendre and  the super-hedging cost is express using the Fenchel bi-conjugate. This allows us to  generalize the result obtained in  \cite{BCL} to the quasi-sure setup. 
Moreover, the super-hedging problem resolution 
leads  endogenously to  a weak quasi-sure no-arbitrage condition  called Absence of Instantaneous Profit (AIP).   An Instantaneous Profit is a very strong condition: one is quasi-sure to make a deterministic and strictly positive profit on a non-polar set.  The AIP condition appears as the minimum requirement in order to get finite prices. 
We show in particular that the AIP is weaker than the quasi-sure no-arbitrage condition of \cite{BN}. 
In a multiperiod model with a time horizon $T$, there is a global Immediate Profit if there is a time $t$ such that it is possible  to super-replicate from a negative cost  paid at time $t$ the claim $0$  at time $T$.  
Thus, a global AIP is defined between times $t$ and $T$ for all $t\in \{0,\ldots,T-1\}.$ This is a major difference with the global no-arbitrage condition which is defined only between $0$ and $T$. Denoting by  $\mathcal{Q}_{t+1}(\o^{t})$ the set of all possible priors for the $t+1$-th period   given the state $\omega^t$ at time $t$ and by $\mathcal{Q}^{T}$ the set of all possible priors between times $0$ and $T$, 
the global no-arbitrage condition requires
$\mathcal{Q}^{T}{\rm -q.s.}$ inequalities when comparing the final wealth  to $0$ while the local no-arbitrage requires $\mathcal{Q}_{t+1}(\o^{t}) {\rm -q.s.}$ ones. The equivalence 
between both notions holds true for $\o^t$ in a $\mathcal{Q}^{t}$-full measure set, see \cite{BN}. 
Here, the novelty is that we have to consider inequalities which hold true for $\mathcal{Q}_{t+1,T}(\o^{t})$ the set of all possible priors for the period between $t+1$ and $T.$ Going from  quasi-sure multiperiod inequalities to quasi-sure  one period inequalities requires technical results which are proved in the appendix. These results allow us to prove the equivalence between the  global AIP and the one-step $\mbox{AIP}_{t}$ for all $t.$ We call a one-step AIP a global AIP in a one period submarket. 
We also show that the one-step AIP is equivalent to a one-period AIP on a full measure set. Defining the global AIP  between times $t$ and $T$ for all $t$ in order 
to obtain the equivalence between the  global AIP and the one-step AIP
 is not due to the quasi-sure framework. This was already the case with a uni-prior model.

The paper is organized as follows. In Section \ref{secappendix}, we introduce the quasi-sure support of a family of $\R^d$-valued functions as well as  the quasi-sure essential supremum of a family of $\R$-valued functions. 
We also provide properties of the quasi-sure essential supremum.  In Section \ref{applifi}, we apply these results to one-period financial problems like  super-replication and  AIP. Section \ref{sec:multi} studies the multiperiod AIP condition. Finally,  the appendix collects some technical results on aggregation of quasi-sure statements.

We finish this introduction with notations and remind of some key results of measure theory. 
For any Polish space $E$ (i.e.  separable topological space that is completely metrizable), we denote by $\mathcal{B} (E)$ its Borel sigma-algebra  and by $\mathfrak{P}(E)$ the set of all probability measures on $(E,\mathcal{B} (E))$. We recall  that $\mathfrak{P}(E)$ endowed with the weak topology is a Polish space (see  \cite[Propositions 7.20 p127, 7.23 p131]{bs}). If $P$ is in $\mathfrak{P}(E)$, $\mathcal{B}_{P}(E)$ will be the completion of $\mathcal{B}(E)$ with respect to $P$ and the universal sigma-algebra is defined by $\mathcal{B}_{c}(E):= \bigcap_{P \in \mathfrak{P}(E)} \mathcal{B}_{P}(E)$. 
In the rest of the paper, we will use the same notation for 
$P$ in $\mathfrak{P}(E)$ and for its (unique) extension on $\mathcal{B}_{P}(E)$. 

For a given $\mathcal{P} \subset  \mathfrak{P}(E)$, 
a set $N \subset E$ is called a $\mathcal{P}$-polar set if there exists some $A \in \mathcal{B}(E)$ such that $N \subset A$ and $P(A)=0$ for all $P \in \mathcal{P}$. 
A property holds true $\mathcal{P}$-quasi-surely (q.s.), if it is true outside a $\mathcal{P}$-polar set and a set is of $\mathcal{P}$-full measure  if its complement is a $\mathcal{P}$-polar set.

Let  $f: E \to F$, where $F$ is an other Polish space and let  $\Ac$ be a sigma-algebra on $E$. We denote by $L^0(F,\Ac)$  the set of $F$-valued and $\Ac$-measurable functions,   i.e.  $f^{-1}(B) \in  \Ac$ for all $B \in \mathcal{B}(F)$. When $\Ac=\mathcal{B}_{c}(E)$ (resp. $\Ac=\mathcal{B}(E)$) $f$ is called 
universally measurable (resp. Borel measurable).

Let  $\Gamma:$ $E \twoheadrightarrow F. $ The notation $\twoheadrightarrow$ stands for set-valued mapping, i.e. $\Gamma : E \to 2^{F},$ where $2^{F}$ is the power set of $F.$ 
Then, $X \in  L^0(F,\Ac)$ is a $\Ac$-measurable selection of $\Gamma$ if $X(\cdot) \in \Gamma(\cdot)$ $  \Pc {\rm -q.s.}$ Moreover, $L_{\Pc}^0(\Gamma,\Ac)$ will denote the family of all  $\Ac$-measurable selections of $\Gamma$: 
\begin{eqnarray}
\label{eqcaselec}
L_{\Pc}^0(\Gamma,\Ac) & := & \{X \in  L^0(F,\Ac): \; X(\cdot) \in \Gamma(\cdot) \; \Pc {\rm -q.s.}\}\\
\label{eqcaselec2}
L_{\Pc}^0(\Gamma,\Ac)  \setminus \{0\}& := & \{X \in  L_{\Pc}^0(\Gamma,\Ac): \; \exists P \in  \Pc, \, P(X\neq 0)>0\}.
\end{eqnarray} 

 An analytic set of $E$ is the continuous image of a Polish space, see  \cite[Theorem 12.24 p447]{Hitch}. We denote by $\mathcal{A}(E)$ the set of analytic sets  of $E$ and  by $\mathcal{C}A(E):=\left\{A \in E:\; E \backslash{A} \in \mathcal{A}(E)\right\}$ the set of co-analytic sets of $E$. Recall that  
the projection of an analytic set is an analytic set  (see  \cite[Proposition 7.39 p165]{bs}), that the countable union, intersection or cartesian product 
 of analytic sets are analytic sets (see \cite[Corollary 7.35.2 p160, Proposition 7.38 p165]{bs}) 
and that (see  \cite[Proposition 7.36 p161, Corollary 7.42.1 p169]{bs})
 \begin{align}
 \label{analyticset}
 \mathcal{B}(E) \subset \mathcal{A}(E) \subset \mathcal{B}_{c}(E).
  \end{align}

Finally,   if $x,y\in\mathbb{R}^d$ then
the concatenation $xy$ stands for their scalar product. The symbol $|\cdot|$ denotes the Euclidean norm
on $\mathbb{R}^d$ (or on $\mathbb{R})$. We also denote by $\bar{\R}=\R\cup\{ \pm \infty\}$ the extended real line.  Recall that $\bar{\R}$ is a Borel space with the weakest topology containing the intervals  $[-\infty,a),$ $(b,\infty]$, $(a,b)$, where $a,b \in \R$, see \cite[p118]{bs}.

\section{Quasi-sure support and quasi-sure essential supremum}
\label{secappendix}

From now,  $\O$ and $\widetilde{\O}$ are two Polish spaces and $\Pc : \Omega \twoheadrightarrow \mathfrak{P}(\widetilde{\O})$ is a nonempty set-valued mapping. The set $\Pc(\o)$ contains all the possible priors on  $\mathfrak{P}(\widetilde{\O})$ given the state $\omega  \in \Omega.$

\subsection{Quasi-sure support of  a vector-valued functions}

The quasi-sure essential supremum will be defined as the supremum of the quasi-sure support, see \eqref{defsupport}. So, we first introduce definitions of measurability for both  set-valued mappings and functions that induce nice measurability properties for both the quasi-sure support and  the quasi-sure  essential supremum. 

Let  $\Oc$ be the class of all open subsets of $\bar{\R}^d.$
\begin{definition} 
Let $\Cc\subset \Oc$. 
\begin{itemize}
\item A function $X:\O \times   \widetilde{\O} \to \bar{\R}^d$ is $\Cc$-analytically measurable if $X^{-1}(C) \in \Ac(\O\times   \widetilde{\O})$ for all $C \in  \Cc$. 
\item If $d=1$ and $\Cc=\{(c,\infty]:\, c \in \R\},$ a $\Cc$-analytically measurable function  is called upper semi-analytic (usa)\footnote{See \cite[Definition 7.21 p177]{bs}.}.
\item A set-valued mapping $J : \Omega \twoheadrightarrow \bar{\R}^d$ is $\Cc$-analytically 
measurable if for all $C\in \Cc,$  
\begin{eqnarray*}
\left\{\o \in \O:\; J(\o) \cap C \neq \emptyset\right\}  \in \mathcal{A}(\O).
\end{eqnarray*}
\item If $d=1$ and $\Cc=\{(c,\infty]: \, c \in \R\},$  a $\Cc$-analytically measurable set-valued mapping is called  
usa-measurable. 
\end{itemize}
\end{definition}
Note that \eqref{analyticset} implies that if $X$ is a Borel measurable function, then $X$ is $\Oc$-analytically measurable and that 
 if $X$ is  usa  then 
 $X$ is $\mathcal{B}_{c}(\O \times   \widetilde{\O})$-measurable. 
 
We now introduce the quasi-sure support of a universally measurable function. 
\begin{definition}
\label{Defsupp} 
Let $X:\O \times   \widetilde{\O} \to \bar{\R}^d$  be a universally measurable function. 
The set-valued mapping  $\supp[\Pc]X : \Omega \twoheadrightarrow \bar{\R}^d$ is called the quasi-sure support of $X$ and is defined by~:
\begin{align*}
\label{defd1}
\supp[\Pc]X(\o) &:=\supp[\Pc(\o)]X(\o,\cdot)\\
&=\bigcap \left\{ F \subset \bar{\R}^d\; \mbox{closed:} \; P(X(\o, \cdot) \in F)=1, \; \forall P \in \Pc(\o)\right\}.
\end{align*}
\end{definition}
We propose in Lemma \ref{Dmeasurability}  an extension of the results obtained by \cite[Lemma 4.3]{BN} and
\cite[Lemma 2.2]{BC16}  on  the measurability of the quasi-sure support of Borel measurable functions  to  $\Oc$-analytically measurable ones. 
This lemma is obtained under the following assumption on ${\rm gph} \,\Pc$ the graph of $\Pc$. This assumption will be used in almost all our results. 
\begin{assumption}
\label{anaone}
The set 
$
{\rm gph} \,\Pc:=\{(\omega,P) \in \Omega \times \mathfrak{P}(\widetilde{\O}):\; P \in \Pc(\omega)\}$ 
is a nonempty analytic set. 
\end{assumption}
\begin{remark}
\label{remgraphana}
Assumption \ref{anaone} has become a standard in the quasi-sure financial literature. Here, we do not assume, as it is usually the case, that  ${\rm gph} \, \Pc$ is a convex set. Apart from Assumption \ref{anaone}, no other assumption is made: $\Pc$ is neither assumed to be dominated by a given probability measure, nor to be weakly compact. This allows for various general definition for $\Pc,$ see \cite{Bart16} or \cite{BC16} for concrete examples of non-dominated settings  where Assumption \ref{anaone} is satisfied as robust binomial trees or discretized $d$-dimensional diffusions. 
\end{remark}
\begin{lemma}
\label{Dmeasurability}
Assume that Assumption \ref{anaone} holds true 
and that  $X:\O \times   \widetilde{\O} \to \bar{\R}^d$ is $\Oc$-analytically measurable,  then $\supp[\Pc]X$ is a  nonempty, closed-valued and $\Oc$-analytically measurable set-valued mapping.  
\end{lemma}
\begin{remark}
\label{simple}
The preceding lemma shows that if $X$ is $\Oc$-analytically measurable (and a fortiori Borel measurable) then $\supp[\Pc]X$ is a nonempty, closed-valued and $\Oc$-analytically measurable set-valued mapping. It  is thus universally measurable in the sense of  \cite[Definition 14.1 p643]{rw},  
 see \eqref{analyticset}.  So, Lemma \ref{Dmeasurability} implies \cite[Lemma 4.3]{BN}. 
\end{remark}
{\sl Proof of Lemma 
\ref{Dmeasurability}.}
As  $X$ is $\Oc$-analytically measurable, it is also universally measurable and $\supp[\Pc]X$ is well-defined. 
 It is clear from its  definition that for all $\o \in \O$,  $\supp[\Pc]X(\o)$ is a {nonempty} and closed subset of $\bar{\R}^d$.
 To prove that $\supp[\Pc]X$ is a $\Oc$-analytically measurable set-valued mapping, we fix some open set $O \in \Oc$. Then, by definition of the support, we get that
\begin{eqnarray*}
\left\{\o \in \O:\; \supp[\Pc]X(\o) \cap O \neq \emptyset\right\}
&=& \left\{\o \in \O:\; \exists P \in \Pc(\o),\, P(X(\o, \cdot)\in O)>0\right\} \\
&= &\kappa^{-1}((0,\infty)),
\end{eqnarray*}
where $\kappa: \; \o \mapsto \kappa(\o):= \sup_{P \in \Pc(\o)}P(X(\o,\cdot)\in O)$. 
Thus, if $\kappa$ is usa, the result is proved. 
We first show that $\O \times \tilde{\O} \times \Pc(\tilde{\O})  \ni (\o,\tilde{\o},P) \mapsto \one_{X(\cdot)\in O}(\o,\tilde{\o})$ is usa. Let $a\in \R$ and $A:=\{(\o,\tilde{\o},P):\; \one_{X(\cdot)\in O}(\o,\tilde{\o}) > a \}$. The set $A$ is empty if $a\geq 1$, equal to $\O \times \tilde{\O} \times \Pc(\tilde{\O})$ if $a<0$. If $a\in [0,1)$, we get that 
$$A=\{(\o,\tilde{\o},P):\; X(\o,\tilde{\o})\in O\}=X^{-1}(O) \times \Pc(\tilde{\O}) \in \mathcal{A}(\O \times \tilde{\O} \times \Pc(\tilde{\O})),$$
using that $X^{-1}(O)  \in \mathcal{A}(\O \times \tilde{\O} )$ by assumption and that 
$\Pc(\tilde{\O}) \in \mathcal{A}(\Pc(\tilde{\O})).$  
Applying \cite[Proposition 7.48 p180]{bs} (as $q(d\tilde{\o}|(\o,P))=P(d\tilde{\o})$ is a Borel stochastic kernel),
$(\o,P) \mapsto P(X(\o,\cdot) \in O)$ is usa. Thus,  as ${\rm gph} \, \Pc$ is assumed to be analytic, \cite[Proposition 7.47 p179]{bs}  implies that $\kappa$ is usa.
$\Box$\\

\subsection{Quasi-sure essential supremum}\label{A1}

When there is a single probability measure prevailing on the state space,  \cite{BCJ}  has incorporated measurability in the definition of the essential supremum (see also \cite{KL} for a family of vector-valued functions).  
 We propose below a quasi-sure version of the  essential supremum.

\begin{theorem}\label{Essup}
Let $I$ be an arbitrary nonempty  index set. Let $\Xc=(X_i)_{i\in I}$ be a family of  universally measurable  functions 
$X_i:\O \times   \widetilde{\O} \to \bar{\R}$ such that $\cup_{i \in I} \supp[\Pc]X_i$ is a nonempty usa-measurable set-valued mapping. 
Then, there exists a  unique usa function $X: \, \O  \to \bar{\R},$ denoted by $\essup[\Pc]\Xc,$ which satisfies the following properties for any fixed $\o \in \O:$
\begin{enumerate}
\item $X(\o) \ge X_i(\o,\cdot)$ $\Pc(\o){\rm -q.s.}$ for all $i\in I.$
\item  Let $y \in \R.$ Then, $y \ge X_i(\o,\cdot)$ $\Pc(\o){\rm -q.s.}$ $\forall i\in I$ if and only if $y\ge X(\o)$.
\end{enumerate}
Moreover, we have that
\begin{eqnarray}
\label{defsupport}
\essup[\Pc] \Xc=\sup \cup_{i \in I} \supp[\Pc]X_i.
\end{eqnarray}
\end{theorem}
Note that condition 2. in Theorem \ref{Essup} also holds true when $y$ is function as $\o$ is fixed:  let $Y:\, \Omega \to \R,$ then, 
$$2\rq{}. \; Y(\o) \ge X_i(\o,\cdot) \, \Pc(\o){\rm -q.s.} \,\forall i\in I \, \mbox{ if and only if } Y(\o)\ge X(\o).$$
One may compare Theorem \ref{Essup} with \cite[Proposition 2.5]{BCL}. This proposition is stated in the uni-prior context and proves that  a $\cal H$-measurable essential supremum exists for  $\cal G$-measurable functions all of them being defined on $\O$, where $\cal H \subset \cal G$ are $\sigma$-algebra on $\O$. Here, we start from functions defined on the product space $\O \times   \widetilde{\O}$ and obtain a quasi-sure essential supremum defined on $\O$. The $\cal G$-measurability of the functions is replaced by usa-measurability on $\O \times   \widetilde{\O}$, while the $\cal H$-measurability of the essential supremum is replaced by  usa-measurability on $\O$. Note that the usa-measurability of the set-valued mapping $\cup_{i \in I} \supp[\Pc]X_i$ is the tailor made condition so that  \eqref{adopt} below holds true. 

{\sl Proof of Theorem \ref{Essup}.}  
We start with unicity. Assume that there exist some $X,\,T: \, \O  \to \bar{\R}$  satisfying 1. and 2.  for any fixed $\o \in \O.$ 
Using 1. for $T,$  we get that $T(\o) \ge X_i(\o,\cdot)$ $\Pc(\o){\rm -q.s.}$ for all $i\in I.$
Thus, 2. for $X$ with $y=T(\o)$  implies that $T(\o) \geq X(\o)$. Using now 1. for $X$ and 2. for $T$ with $y=X(\o),$  we get that $X(\o) \geq T(\o)$ and the unicity is proved. 
We now prove existence. Assume first that $(X_i)_{i \in I}$ takes values in $[0,1].$ 
Fix some $\o \in \O.$ Let $y \in \R$. Then,  
\begin{eqnarray*}
y \ge X_i(\o,\cdot) \, \Pc(\o){\rm -q.s.} & \Leftrightarrow & \forall P \in \Pc(\o),\; P(X_i(\o,\cdot) \leq y)=1\\
 & \Leftrightarrow & \supp[\Pc]X_i(\o) \subset ]-\infty, y].
\end{eqnarray*}
Thus, $y \ge X_i(\o,\cdot)\, \Pc(\o){\rm -q.s.}$ for all $ i\in I$ if and only if $\cup_{i \in I}\supp[\Pc]X_i(\o) \subset ]-\infty, y]$.
Let \begin{eqnarray}
\label{sup01}X(\o):=\sup\{x \in [0,1]: \; x \in \cup_{i \in I}\supp[\Pc]X_i(\o)\}.
\end{eqnarray} 
We prove that $X$ satisfies 1. and 2.
Indeed,  
$y \ge X_i(\o,\cdot)$ $\Pc(\o){\rm -q.s.}$ for all $i\in I$ if and only if  
$y \geq X(\o)$ and this shows 1. Moreover, as for every $i\in I$ 
$P(X_i (\o,\cdot) \in \cup_{i \in I}\supp[\Pc]X_i(\o))=1$ for all $P \in \Pc(\o)$ (see \cite[Lemma 4.2]{BN}),  we get that
$X(\o) \ge X_i(\o,\cdot)$  $\Pc(\o){\rm -q.s.}$ and 2. is proved. 

We show now that 
$X$ is usa. Fix some $c\in \R.$ As $\cup_{i \in I} \supp[\Pc]X_i$ is a usa-measurable set-valued mapping, we get that
\begin{eqnarray}
\label{adopt}
\{ X> c\} & = & \{\o \in \O: \;   \cup_{i \in I} \supp[\Pc]X_i(\o) \cap (c,+\infty) \neq \emptyset \}\in \Ac(\O). 
\end{eqnarray}

Assume now that $(X_i)_{i \in I}$ takes values in $\bar{\R}$. We consider an increasing homeomorphism $\Psi: \bar{\R} \to [0,1]$ which inverse is also increasing and set 
$$\essup[\Pc]\Xc :=\Psi^{-1} \left(\essup[\Pc] \{\Psi(X_i): \; i \in I\} \right).$$ 
It is easy to see that $\essup[\Pc]\Xc$ verifies $1.$ and $2.$ Moreover, the definition of $\essup[\Pc]\Xc$ does not depend from the choice of $\Psi$ by unicity of $\essup[\Pc]\Xc$. Then,  we obtain that 
\begin{eqnarray*}
\essup[\Pc]\Xc(\o) & = & \Psi^{-1} \left(\sup \cup_{i \in I} \supp[\Pc] \Psi(X_i) (\o)\right)\\
& = & \sup \Psi^{-1} \left(\cup_{i \in I} \supp[\Pc] \Psi(X_i) (\o)\right)\\
& = & \sup \cup_{i \in I} \supp[\Pc] X_i(\o). 
\end{eqnarray*}
The first equality follows from \eqref{sup01}, the second one from the fact that $\Psi^{-1} $ is non decreasing. For the last one we remark that as  $\Psi$ and $\Psi^{-1} $ are  continuous $\Psi^{-1} \left(\supp[\Pc] \Psi(X_i) (\o)\right)=\supp[\Pc] X_i(\o). $ Indeed, let $\Oc_1$ be the class of all open subsets of $\bar{\R}.$
\begin{eqnarray*}
x \notin \Psi^{-1} \left(\supp[\Pc] \Psi(X_i) (\o)\right) & \Leftrightarrow  & \Psi(x) \notin \supp[\Pc] \Psi(X_i) (\o) \\
& \Leftrightarrow  & \exists O \in \Oc_1 \mbox{ s.t. } \psi(x) \in O \mbox{ and } \\
& & P(\Psi(X_i(\o,\cdot)) \in O) =0, \; \forall P \in \Pc(\o)  \\
& \Leftrightarrow  & \exists O \in \Oc_1 \mbox{ s.t. } x \in \Psi^{-1}(O) \mbox{ and } \\
& & P(X_i(\o,\cdot) \in \psi^{-1}(O)) =0, \; \forall P \in \Pc(\o)  \\
& \Leftrightarrow  & x \notin\supp[\Pc] X_i(\o). 
\end{eqnarray*}
Thus, we have proved \eqref{defsupport} and the proof is complete. 
$\Box$\\

The following lemma provides tractable conditions for the existence of the quasi-sure essential supremum. 
\begin{lemma}
Let $I$ be a nonempty countable set and let $\Xc=(X_i)_{i\in I}$ be a family of usa functions
$X_i:\O \times   \widetilde{\O} \to \bar{\R}.$ Assume that $\Pc$ satisfies Assumption \ref{anaone}. Then,  $\cup_{i \in I} \supp[\Pc]X_i$ is a nonempty usa-measurable set-valued mapping. 
\end{lemma}
{\sl Proof.} 
As $X_i$ is usa and Assumption \ref{anaone} holds true, Lemma \ref{Dmeasurability}  shows that  $\supp[\Pc]X_i$ is a nonempty usa-measurable set-valued mapping.
Let $c \in \R$. Then 
\begin{eqnarray*}
\scalemath{0.95}{
\left\{\o \in \O:\; \cup_{i \in I} \supp[\Pc]X_i(\o) \cap (c,\infty] \neq \emptyset\right\}=
\cup_{i \in I} \left\{\o \in \O:\;  \supp[\Pc]X_i(\o) \cap (c,\infty] \neq \emptyset\right\}
}
\end{eqnarray*}
\hspace*{-0.1cm} belongs to $\mathcal{A}(\O)$ as $I$ is countable and the claim is proved. 
$\Box$\\
For financial applications $\Pc$ satisfies Assumption \ref{anaone} as mention in  Remark \ref{remgraphana} and we will consider the quasi-sure essential supremum of a single function $X$ which will be the sum of the payoff of the contingent claim and of the cost of the super-hedging strategy. In the quasi-sure literature the contingent claim is usually  assumed to be usa and the cost function  to be Borel measurable so that $X$ is in turn usa.\\
Assume that  $X:\O \times   \widetilde{\O} \to \bar \R$ is a usa function and that $\Pc$ satisfies Assumption \ref{anaone}, 
 the  random interval ${[}_{\Pc}X(\o,\cdot),+\infty)$ is characterised as follows 
\begin{eqnarray}
\label{alea}
{[}_{\Pc}X(\o,\cdot),+\infty):=\left\{x\in \R: \;    x  \ge X(\o,\cdot) \, \Pc(\o){\rm -q.s.}\right\}=[\essup[\Pc] X(\o),\infty).
\end{eqnarray}
Note that the preceding random interval is empty if $\essup[\Pc] X(\o)=\infty$.

\subsection{Properties of the quasi-sure essential supremum}\label{A1}

This section collects various applications of Theorem \ref{Essup} which are of own interest and which will serve for solving financial problems such as super-replication or the characterization of the Absence of Instantaneous Profit condition. 

This first proposition shows that under suitable conditions the quasi-sure essential supremum of $h(X)$ equals the regular supremum of $h$ taken over $\supp[\Pc]X$. This proposition will be used to provide a dual representation of the super-hedging prices from which follows a bidual representation of the minimum of these super-hedging prices (called the super-hedging cost). 
\begin{proposition}\label{lemma-essup-h(X)}
Assume that Assumption \ref{anaone} holds true and let $X:\O \times   \widetilde{\O} \to \R^d$ be   a Borel measurable function.   
Let $h:$ $\Omega \times \R^d \to \R \cup \{+\infty\}$ be a usa function. Fix   $\o \in \O$ and assume that  $h(\o,\cdot)$ is lower semi-continuous (lsc).  
Then, we get that 
\begin{eqnarray}
\label{belequa}
\essup[\Pc] h(X)(\o)=\sup_{x\in\supp[\Pc]X(\o)} h(\o,x),
\end{eqnarray}
where $h(X):(\o,\tilde \o) \mapsto h(X)(\o,\tilde \o):=h(\omega,X(\omega,\tilde \o)).$ 
\end{proposition}
As $h(X)$ is usa (see \cite[Lemma 7.30 p177]{bs}), the quasi-sure essential supremum of $h(X)$ exists. Thus,  \eqref{defsupport} and  \eqref{belequa} imply that 
\begin{eqnarray*}
\sup \supp[\Pc] h(X)(\o)=\sup_{x\in\supp[\Pc]X(\o)} h(\o,x). 
\end{eqnarray*}
Proposition \ref{lemma-essup-h(X)} generalizes \cite[Proposition 2.7]{BCL} to the quasi-sure setting. The proof is indeed easily adapted since the uni-prior and the multi-prior supports are characterized as follows (see \cite[Theorem 12.14 p442]{Hitch}, \cite[Lemma 4.2]{BN} and \cite[Lemma 5.2]{BC19}). Let $B\left(x,{1}/{k}\right)$ be the open ball of center $x$ and radius $1/k$. 
Let $P \in \mathfrak{P}(\widetilde{\O}).$ Then, $x \in \supp[P]X(\o)$ if and only if for all $k \geq 1,$ 
$P\left(X(\o,.) \in B\left(x,{1}/{k}\right)\right)>0.$ 
Now,  $x \in \supp[\Pc]X(\o)$ if and only if for all $k \geq 1$, there exists  $P^{k} \in \mathcal{P}(\o)$ such that  $P^{k}\left( X(\o,\cdot) \in B\left(x,{1}/{k}\right)\right)>0$.\\

\noindent {\sl Proof of Proposition \ref{lemma-essup-h(X)}.}
As $P(X(\o,\cdot) \in \supp[\Pc]X(\o))=1$ for all $P \in \Pc(\o)$ (see \cite[Lemma 4.2]{BN}), we have that
$\sup_{x\in \supp[\Pc]X(\o)} h(\o,x) \geq h(\o,X(\o,\cdot))$  $\Pc(\o){\rm -q.s.}$  
As $X$ is Borel measurable and $h$ is usa, $h(X)$ is usa (see \cite[Lemma 7.30 p177]{bs}). Theorem \ref{Essup} implies that 
$\essup[\Pc] h(X)$ exists and  that  
$$\sup_{x\in \supp[\Pc]X(\o)} h(\o,x) \geq\essup[\Pc] h(X)(\o).$$ 
As  $X$ is  Borel measurable, $\supp[\Pc]X$  is a closed-valued and universally measurable  set-valued mapping (see Remark \ref{simple}) and there exists a Castaing representation  $(\gamma_n)_n$ of $\supp[\Pc]X,$ see  \cite[Theorem 14.5 p646]{rw}. So,  Lemma 4.1 of \cite{BCL} implies that
$$\sup_{x\in \supp[\Pc]X(\o)} h(\o,x)=\sup_n h(\o,\gamma_n(\o)).$$ Fix some $n \in \N$ and  $ k>0$. 
Then, as $\gamma_n(\o) \in \supp[\Pc]X(\o)$ there exists some $P^k \in\Pc(\o)$ such that $P^k(X(\o,\cdot)\in  B(\gamma_n(\o),1/k)) >0.$ 
By definition of the quasi-sure essential supremum, we  get that 
$\essup[\Pc] h(X)(\o)\geq h(\o,X(\o,\cdot))$  $\Pc(\o){\rm -q.s.}$ and thus $P^k$-a.s. There exists a $P^k$-full measure subset of $\widetilde{\O}_k \subset \widetilde{\O}$ such that 
\begin{eqnarray*}
\essup[\Pc] h(X) (\o) & \geq &   \frac{\int_{\widetilde{\O}_{k}}1_{B(\gamma_n(\o),1/k)}(X(\o,\cdot))h(\o,X(\o,\cdot)) dP^{k}}{\int_{\widetilde{\O}_{k}}1_{B(\gamma_n(\o),1/k)}(X(\o,\cdot)) dP^{k}}\\
&=&
\frac{\int_{\R^d}1_{B(\gamma_n(\o),1/k)}(x )h(\o,x)  P^k_{X(\o,\cdot)}(dx)}{\int_{\R^d } 1_{B(\gamma_n(\o),1/k)}(x) P^k_{X(\o,\cdot)}(dx)}\\
& \geq   &   \frac{\int_{\R^d } \inf_{y \in B(\gamma_n(\o),1/k)}
h(\o,y) 1_{B(\gamma_n(\o),1/k)}(x ) P^k_{X(\o,\cdot)}(dx)
}{\int_{\R^d } 1_{B(\gamma_n(\o),1/k)}(x) P^k_{X(\o,\cdot)}(dx)}\\
& = & \inf_{y \in B(\gamma_n(\o),1/k)}h(\o,y),
\end{eqnarray*}
where $P^k_{X(\o,\cdot)}$ denotes the law of $X(\o,\cdot)$ under $P^k.$ 
As $h(\o,\cdot)$ is lsc (recall \cite[Definition 1.5, equation 1(2) p8]{rw}), we have that
$$\lim_{k \to \infty} \inf_{y \in B(\gamma_n(\o),1/k)} h(\o,y)=\liminf_{x \to \gamma_n(\o)}h(\o,x)=h(\o,\gamma_n(\o)).$$
It follows that
$\essup[\Pc] h(X) (\o) \geq h(\o,\gamma_n(\o))$. Taking the supremum over all $n$, we get that
$$\essup[\Pc] h(X) (\o) \geq \sup_n h(\o,\gamma_n(\o))=\sup_{x\in \supp[\Pc]X(\o)} h(\o,x).$$ 
$\Box$


We now propose some other applications of Theorem \ref{Essup}. 
First, for a function $\Phi : \Omega \times   \widetilde{\O} \times \R^d \to \R \cup \{+\infty\},$ we study how the assumptions postulated on $\Phi$ carry over to the essential supremum of $(\o,\widetilde{\o}) \mapsto \Phi( \o,\widetilde{\o},\theta)$. We will apply these results when $\Phi( \o,\widetilde{\o},\theta)$ equals the cost of a strategy $\theta$ plus the payoff of some contingent claim, both in the state $( \o,\widetilde{\o}).$ 
We first introduce $ {\rm epi}_{\Pc} \Phi$  the epigraphical set-valued mapping  of $\Phi.$
\begin{definition}\label{zorrettedef}
Let $ {\rm epi}_{\Pc} \Phi: \Omega \twoheadrightarrow \R^d \times \R$ be defined as  follows 
 \begin{eqnarray}
\label{laballephi}
{\rm epi}_{\Pc} \Phi (\o) &:= & \{(\theta,\a) \in   \R^{d}  \times \R: \;    \a \ge \Phi (\o,\cdot,\theta) \, \Pc(\o) {\rm -q.s.}\}. 
\end{eqnarray}
\end{definition}
Next, we will say that a property holds true $\Pc(\o)$-q.s. if it holds true on a $\Pc(\o)$-full measure set in the following sense.  The function  $\theta \mapsto \Phi( \o,\cdot,\theta)$ is 
\begin{itemize}
\item  $\Pc(\o)$-q.s. convex if  $\forall \l \in [0,1],\theta,\theta\rq{} \in \mathbb{R}^d,$ 
$$ \l \Phi( \o,\cdot,\theta) + (1-\lambda) \Phi( \o,\cdot,\theta\rq{}) \geq \Phi( \o,\cdot,\l \theta +(1- \l) \theta\rq{})\, \Pc(\o) {\rm -q.s.}$$
\item $\Pc(\o)$-q.s. lsc if  $\forall \alpha \in \R$, 
$\{\theta \in \R^d: \; \a \geq \Phi( \o,\cdot,\theta)   \, \Pc(\o) {\rm -q.s.}\}$ is closed.\\
\item $\Pc(\o)$-q.s. level coercive, if 
 $\liminf_{|\theta| \to \infty} \frac{\Phi( \o,\cdot,\theta) }{|\theta|}>0  \; \Pc(\o) {\rm -q.s.}$
\end{itemize}
We now introduce conditions on $\Phi$  which will allow us to obtain some measurability properties. We recall that for any set-valued mapping $\Kc:\O \twoheadrightarrow \R^k$, the domain of  $\Kc$ is defined by   $${\rm dom}  \, \Kc :=\{\o \in \Omega: \;  \Kc(\o)\cap \R^k \neq \emptyset\}.$$ 

\begin{assumption}
\label{hypophi2}
Let  $\Phi : \Omega \times   \widetilde{\O} \times \R^d \to \R \cup \{+\infty\}.$  
For all $\o \in {\rm dom} \, {\rm epi}_{\Pc} \Phi $, $\theta \in \R^d$ and $\e>0$, there exists some $\hat \theta \in \Q^d$ such that 
\begin{eqnarray}
\label{lighthypo2}
\Phi( \o,\cdot,\hat \theta) - \Phi( \o,\cdot, \theta) \leq \e  \; \Pc(\o) {\rm -q.s.}
\end{eqnarray}
\end{assumption}
\begin{assumption}
\label{hypophi3}
Let  $\Phi : \Omega \times   \widetilde{\O} \times \R^d \to \R \cup \{+\infty\}.$  For any $\o \in {\rm dom} \, {\rm epi}_{\Pc} \Phi $, there exists some $\eta>0$ such that for all $\e>0$, $\theta \in \R^d$, there exists some $\hat \theta \in \Q^d$ such that 
$|\theta - \hat \theta |<\eta \e$ and 
\begin{eqnarray}
\label{lighthypo3}
\Phi( \o,\cdot,\hat \theta) - \Phi( \o,\cdot, \theta) \leq \e  \; \Pc(\o) {\rm -q.s.}
\end{eqnarray}
\end{assumption}
Of course Assumption \ref{hypophi3} implies Assumption \ref{hypophi2}. Assumption \ref{hypophi3} is implied (but is much weaker) than some kind of partial uniform continuity: 
 for all $\e>0,$ there exists some $\eta>0$ such that for all $\theta, \, \theta' \in \R^d$ satisfying
$|\theta - \theta' |<\eta$, we have that $\Phi( \o,\cdot,\theta') - \Phi( \o,\cdot, \theta) \leq \e  \; \Pc(\o) {\rm -q.s.}$\\
Assumption \ref{hypophi3} is automatically satisfied in frictionless financial applications, see Lemma \ref{unifcont}. 

The following proposition collects some important  properties on the quasi-sure essential supremum of $\Phi( \cdot,\cdot,\theta)$. In particular, it gives conditions for the essential supremum to be jointly measurable, which is a nontrivial result. All the proofs are based on Theorem \ref{Essup}.
\begin{proposition}\label{zorro1}
Assume that Assumption \ref{anaone}  holds true.  Let  $\Phi : \Omega \times   \widetilde{\O} \times \R^d \to \R \cup \{+\infty\}.$ Assume that for all $\theta \in \R^d$, $\Phi( \cdot,\cdot,\theta):(\o,\widetilde{\o}) \mapsto \Phi( \o,\widetilde{\o},\theta)$ is usa.   
Let $\f:\O \times \R^d \to \R \cup \{+\infty\}$ be  the quasi-sure essential supremum of $\Phi( \cdot,\cdot,\theta):$
$$\f(\o,\theta):= \essup[\Pc] \Phi( \cdot,\cdot,\theta) (\o).$$
Then, $\o \mapsto \f(\o,\theta)$ usa. \\
If for all $\o \in \Omega,$ $\theta \mapsto \Phi( \o,\cdot,\theta)$ is $\Pc(\o)$-q.s. convex (resp. lsc,  level coercive), then 
$\theta \mapsto \f(\o,\theta)$ is convex (resp. lsc, level coercive). \\
If Assumption \ref{hypophi3} holds true and if for all $\o \in \Omega,$ $\theta \mapsto \Phi( \o,\cdot,\theta)$ is $\Pc(\o)$-q.s. lsc, then 
$\f $ is   $\Bc_c(\O) \otimes \Bc(\R^d)$-measurable.
\end{proposition}
{\sl Proof.} \\
{\it The function $ \f(\cdot,\theta)$ is usa.} \\
This is a direct application of Theorem \ref{Essup} which shows that for all $\theta \in \R^d,$
$\o \mapsto \f(\o,\theta)=\essup[\Pc]\Phi( \cdot,\cdot,\theta) (\o)$ exists, is unique and usa.\\ 
{\it If for all $\o \in \Omega,$ $\theta \mapsto \Phi( \o,\cdot,\theta)$ is $\Pc(\o)$-q.s. convex (resp. lsc,  level coercive), then 
$\theta \mapsto \f(\o,\theta)$ is convex (resp. lsc, level coercive).}\\
Fix $\o \in \O$. Assume that $\theta \mapsto \Phi( \o,\cdot,\theta)$ is $\Pc(\o)$-q.s. convex. Let $\l \in [0,1]$ and $\theta,\theta\rq{} \in \R^d$.  By 1. in Theorem \ref{Essup} and the convexity property, we get that $\Pc(\o)$ -q.s.
\begin{eqnarray*} 
\l \f(\o,\theta)  + (1-\l) \f(\o,\theta\rq{}) & \geq &  \l \Phi( \o,\cdot,\theta)  + (1-\l) \Phi( \o,\cdot,\theta\rq{})
 \\     & \geq  & \Phi( \o,\cdot,\l \theta +(1- \l) \theta\rq{}).
\end{eqnarray*}
Thus, 2. in Theorem \ref{Essup} shows that 
\begin{eqnarray*} 
\l \f(\o,\theta)  + (1-\l) \f(\o,\theta\rq{}) & \geq &   \essup[\Pc]\Phi( \cdot,\cdot,\l \theta +(1-\l) \theta\rq{}) (\o)\\
& = &  \f(\o,\l \theta + (1-\l) \theta\rq{})
\end{eqnarray*}
and $\theta \mapsto \f(\o,\theta)$ is convex. \\
Assume now that $\theta \mapsto \Phi( \o,\cdot,\theta)$ is $\Pc(\o)$-q.s. lsc. 
We prove that for every $\a \in \R,$ 
$\Theta:=\{\theta \in \R^d: \; \a \geq  \f( \o, \theta)\}$ 
is closed. Let $(\theta_n)_n \subset \Theta$ that converges to some $\theta^*$. 
Then, for all $n$, 1. in Theorem \ref{Essup} shows that 
\begin{eqnarray*} 
\a \geq  \f( \o, \theta_n)=\essup[\Pc]\Phi( \cdot,\cdot,\theta_n) (\o) \geq \Phi( \o,\cdot,\theta_n)  \, \Pc(\o) {\rm -q.s.} 
\end{eqnarray*}
So, $(\theta_n)_n \subset \{\theta \in \R^d: \;  \a \geq \Phi( \o,\cdot,\theta)  \, \Pc(\o) {\rm -q.s.}\},$ which is closed by assumption and thus
$
\a  \geq \Phi( \o,\cdot,\theta^*) \, \Pc(\o) {\rm -q.s.}$ 
So, 2. in  Theorem \ref{Essup} shows that $$\a \geq \essup[\Pc]\Phi( \cdot,\cdot,\theta^*) (\o)=\f(\o,\theta^*)$$ and    $\theta^* \in \Theta.$ Thus,  $\theta \mapsto \f(\o,\theta)$ is lsc.\\
Assume now that $\theta \mapsto \Phi( \o,\cdot,\theta)$ is $\Pc(\o)$-q.s. level coercive. 
Let $(\theta_n) \subset \R^d$ such that $|\theta_n| \to \infty$ when $n$ goes to infinity. Then, 
1. in Theorem \ref{Essup} shows that for all $n\geq 1$ 
$$ \frac{\f(\o,\theta_n)  }{|\theta_n|} \geq  \frac{\Phi( \o,\cdot,\theta_n) }{|\theta_n|} \; \Pc(\o) {\rm -q.s.}$$
We also obtain that for all $P \in \Pc(\o)$ 
 $$ P \left( \cap_{n \geq 1} \left\{\frac{\f(\o,\theta_n)  }{|\theta_n|} \geq  \frac{\Phi( \o,\cdot,\theta_n) }{|\theta_n|}\right\}\right)=1.$$
Thus, the level coercive property of $\Phi$ implies that 
$$ \liminf_{n \to \infty}\frac{\f(\o,\theta_n)  }{|\theta_n|} \geq  \liminf_{n\to \infty} \frac{\Phi( \o,\cdot,\theta_n) }{|\theta_n|}>0  \; \Pc(\o) {\rm -q.s.}$$
and $\theta \mapsto \f(\o,\theta)$ is level coercive.

\noindent {\it The function $\f$ is   $\Bc_c(\O) \otimes \Bc(\R^d)$-measurable.}\\ 
We will prove in Proposition \ref{zorrette} below that $ {\rm epi}_{\Pc} \Phi$ (see \eqref{laballephi})  the epigraphical set-valued mapping  of $\Phi$ is closed-valued and $\Bc_c(\Omega)$-measurable. Using  \cite[Theorem 14.8 p648]{rw}, this implies that 
${\rm gph} \,  {\rm epi}_{\Pc}\Phi \in  \Bc_c(\Omega) \otimes \Bc(\R^d) \otimes \Bc(\R) .$    Thus, 
for all $\alpha \in \R$, the section $S_{\a}$ of  ${\rm gph} \,  {\rm epi}_{\Pc} \Phi $ belongs to $\Bc_c(\Omega) \otimes \Bc(\R^d),$ see \cite[Lemma 4.46 p151]{Hitch}.
Using \eqref{painepi} below, this section $S_{\a}$ can be written as follows
\begin{eqnarray*}
S_{\a} & := & \{(\o, \theta) \in \Omega \times \mathbb{R}^d: \;  (\o,\theta,\alpha) \in {\rm gph} \,  {\rm epi}_{\Pc} \Phi  \}\\
 & = & \{(\o, \theta) \in \Omega \times \mathbb{R}^d: \;  (\theta,\alpha) \in  {\rm epi}_{\Pc} \Phi (\o) \}= \{(\o, \theta) \in \Omega \times \mathbb{R}^d: \; \a \geq \f(\o,\theta)\}
 \end{eqnarray*}
 which achieves the proof.  
$\Box$\\

We exhibit in the next proposition a class of $\Phi$ that are epi-measurable, i.e. such that ${\rm epi}_{\Pc} \Phi$ the epigraphical set-valued mapping of $\Phi$  (see \eqref{laballephi}) is $\Bc_c(\Omega)$-measurable  and also graph-measurable. This  nontrivial result was used  in the preceding proof. 
\begin{proposition}\label{zorrette}
Assume that Assumption \ref{anaone}  holds true.  Let  $\Phi : \Omega \times   \widetilde{\O} \times \R^d \to \R \cup \{+\infty\}$ be a function satisfying Assumption \ref{hypophi3}. Assume also that for all $\theta \in \R^d$, $\Phi( \cdot,\cdot,\theta):(\o,\widetilde{\o}) \mapsto \Phi( \o,\widetilde{\o},\theta)$ is usa and that $\theta \mapsto \Phi( \o,\cdot,\theta)$ is $\Pc(\o)$-q.s. lsc  for all $\o \in \Omega.$ 
Then, ${\rm epi}_{\Pc} \Phi$ is a closed-valued and $\Bc_c(\Omega)$-measurable  set valued mapping (see \cite[Definition 14.1]{rw} p643), ${\rm dom} \,{\rm epi}_{\Pc} \Phi$ is a co-analytic set and 
 \begin{eqnarray}
\label{painepi} 
{\rm epi}_{\Pc} \Phi (\o) &= {\rm epi} \, \f (\o,\cdot) =\{(\theta,\alpha) \in \mathbb{R}^d \times \R : \; \a \geq \f( \o, \theta)  \}. 
\end{eqnarray}
\end{proposition}
{\sl Proof.} 
As $\Phi (\cdot,\cdot,\theta)$ is usa, $\f$ is well defined and \eqref{painepi} follows immediately from the definition of the epigraph in  \eqref{laballephi} and from Theorem \ref{Essup}. 

Fix $\o \in \O$. 
As $\theta \mapsto \Phi( \o,\cdot,\theta)$ is $\Pc(\o)$-q.s. lsc, ${\rm epi}_{\Pc} \Phi (\o)$ is closed. 
Thus, to prove that  ${\rm epi}_{\Pc} \Phi$ is a $\Bc_c(\Omega)$-measurable set-valued mapping, we can apply \cite[Theorem 14.3 p644]{rw} and show that for all $x\in \R^{d+1}$ and $\e>0$
\begin{eqnarray*}
A_{\e,x}& := & \{\o \in \Omega:\; {\rm epi}_{\Pc}  \Phi(\o) \cap B(x,\e) \neq \emptyset\}\in \Bc_c(\Omega),
\end{eqnarray*}
where $B(x,\e)$ is the open ball of center $x$ and radius $\e$. Remark that 
\begin{eqnarray*}
A_{\e,x}& = & \{\o \in {\rm dom} \,{\rm epi}_{\Pc} \Phi:\; {\rm epi}_{\Pc}  \Phi(\o)\cap B(x,\e) \neq \emptyset\}.
\end{eqnarray*}
We will show that $A_{\e,x}$ is a co-analytic set and thus also a universally measurable set. 
Assume for a moment that we have proved that for all $\o \in {\rm dom} \,{\rm epi}_{\Pc} \Phi$
\begin{eqnarray}
\label{cire2}{\rm epi}_{\Pc} \Phi(\o)= \overline{\{(\theta,\alpha) \in \mathbb{Q}^d \times  \mathbb{Q} : \; \a \geq \f( \o, \theta)   \}}
\end{eqnarray}
and that  ${\rm dom} \, {\rm epi}_{\Pc} \Phi  \in \Cc A(\Omega).$
Then,
\begin{eqnarray*}
A_{\e,x} & =& 
\{\o \in {\rm dom} \, {\rm epi}_{\Pc} \Phi:\; \{(\theta,\alpha) \in \mathbb{Q}^d \times  \mathbb{Q} : \; \a \geq  \f( \o, \theta)   \} \cap B(x,\e) \neq \emptyset\} \\
& =& {\rm dom} \, {\rm epi}_{\Pc} \Phi \cap \cup_{(\theta,\alpha) \in (\mathbb{Q}^d \times  \mathbb{Q} )\cap B(x,\e)} \{\o \in \O: \;  \a \geq \f( \o, \theta) \}. 
\end{eqnarray*}
As $\o \mapsto \f( \o, \theta)$ is usa, $\{\o \in \O: \;  \a \geq \f( \o, \theta)  \} \in \Cc A (\O).$ So,  $A_{\e,x} \in  \Cc A (\O) \subset \Bc_c(\Omega)$ and the proof is complete.

Now, we show \eqref{cire2}. Fix $\o \in {\rm dom} \,{\rm epi}_{\Pc} \Phi$. 
As ${\rm epi}_{\Pc} \Phi (\o)$ is closed, the reverse inclusion follows from \eqref{painepi}. Let $(\theta,\alpha) \in {\rm epi}_{\Pc}\Phi (\o).$ Fix some $n\in \N^*.$ There exists some $\a_n \in \Q$ such that 
$\a_n \leq \a \leq \a_n + \frac1{4n}.$
We apply Assumption \ref{hypophi3} and choose $\e \in \mathbb{Q}$ such that $0<\e \leq \frac{1}{2n(\eta+1)}$ and some 
 $\hat \theta \in \Q^d$ such that 
$|\theta - \hat \theta |<\eta \e$ and 
$$\Phi( \o,\cdot,\hat \theta) \leq  \Phi( \o,\cdot, \theta) + \e \leq \a +\e \leq  \a_n + \frac1{4n} +\e=:\hat \a \; \Pc(\o) {\rm -q.s.}$$
Thus, $(\hat \theta, \hat \alpha) \in {\rm epi}_{\Pc} \Phi (\o) \cap \Q^{d+1}.$ Moreover, 
\begin{eqnarray*}|(\hat \theta, \hat \alpha) -(\theta,  \alpha)| & \leq  & |\hat \theta- \theta| +|\hat \a- \a| \leq \eta \e + \frac1{4n} + \frac1{4n} +\e  \leq \frac1{n},
\end{eqnarray*}
and the proof of \eqref{cire2} is complete. \\
Now, we show that $\rm {dom \, epi}_{\Pc} \Phi$ is a co-analytic set.  
The proof is very similar to the proof of \eqref{cire2} and is reported for sake of completeness. 
Note that for this, we only need Assumption \ref{hypophi2}. 
\begin{eqnarray}
\nonumber
{\rm dom} \, {\rm epi}_{\Pc} \Phi  & := & \{ \o \in \Omega:\; {\rm epi}_{\Pc}  \Phi(\o) \cap ( \R^d \times \R) \neq \emptyset\}\\
\nonumber
\nonumber
 & = & \{ \o \in \Omega:\; \exists (\theta,\alpha) \in \mathbb{R}^d \times \R : \;  \a \geq \Phi( \o,\cdot, \theta)  \;   \, \Pc(\o) {\rm -q.s.} \}\\
 \label{chat}
 & = & \{ \o \in \Omega:\; \exists (\theta,\alpha) \in \mathbb{Q}^d \times \mathbb{Q} : \; \a \geq \Phi( \o,\cdot, \theta)  \;   \, \Pc(\o) {\rm -q.s.} \}\\
 \nonumber
 &= & \cup_{(\theta,\alpha) \in (\mathbb{Q}^d \times  \mathbb{Q} )} \{ \o \in \Omega:\;  \a \geq \Phi( \o,\cdot, \theta)   \;   \, \Pc(\o) {\rm -q.s.} \}\\
  \nonumber
 &= & \cup_{(\theta,\alpha) \in (\mathbb{Q}^d \times  \mathbb{Q} )} \{ \o \in \Omega:\;  \a \geq \f( \o, \theta) \} \in \Cc A (\Omega),
\end{eqnarray}
as $\o \mapsto \f( \o, \theta)$ is usa. 
Equality \ref{chat} is proved as follows: let $\o \in \O$ be such that there exists  $ (\theta,\alpha) \in \mathbb{R}^d \times \R $ such that $\Phi( \o,\cdot, \theta) \leq \a  \;   \, \Pc(\o) {\rm -q.s.}$ Let $\e \in \Q$. We choose some some $\hat \a \in \Q$ such that 
$\hat \a \leq \a \leq \hat \a  + \frac{\e}{2}.$ Using Assumption \ref{hypophi2} (here $\o \in {\rm dom} \, {\rm epi}_{\Pc} \Phi$), we find some $\hat{\theta} \in \Q^d$ such that 
\begin{align*}
\Phi( \o,\cdot,\hat \theta) - \Phi( \o,\cdot, \theta) \leq \frac{\e}2  \; \Pc(\o) {\rm -q.s.}
\end{align*}
Then, 
\begin{align*}
\Phi( \o,\cdot,\hat \theta) = \Phi( \o,\cdot,\hat \theta) - \Phi( \o,\cdot, \theta) + \Phi( \o,\cdot, \theta)  \leq \frac{\e}2 + \a \leq \hat \a  + \e \; \Pc(\o) {\rm -q.s.}
\end{align*}
As $(\hat \theta, \hat \a  + \e)  \in \mathbb{Q}^d \times \mathbb{Q},$ the equality is proved in \eqref{chat}.  
$\Box$\\

We introduce now the projection on $\R$ of the epigraph and the infimum of this projection. This  set will be the set of super-hedging prices in financial applications while the infimum will be called the super-hedging cost. 
\begin{definition}
\label{defpphi}
Let $\Phi : \Omega \times   \widetilde{\O} \times \R^d \to \R\cup\{ + \infty\}.$ 
Let $\Pi(\Phi): \Omega \twoheadrightarrow \R$  be the set-valued mapping defined as follows 
 \begin{eqnarray}
\nonumber
\Pi(\Phi)(\o)& := & {\rm proj}_{\R} {\rm epi}_{\Pc} \Phi (\o)  \\
\label{laballePhi}
&  = & \{x\in  \R: \;   \exists\, \theta\in\R^d,\;   x \ge \Phi (\o,\cdot,\theta) \, \Pc(\o) {\rm -q.s.}\}.
\end{eqnarray}
Let  $\pi(\Phi)(\o):=\inf \Pi(\Phi)(\o),$
with the convention (which will be used until the end of the paper) that $\inf \emptyset = +\infty$.
\end{definition}
Note that $\theta$ in the financial applications plays the role of the super-hedging strategy.  
We provide now some characterization of $\Pi(\Phi)$ and $\pi(\Phi).$ We also prove  some nontrivial measurable selection result on the $\theta$ appearing in $\Pi(\Phi)$ which will be of great importance in the multiperiod setting when proving Proposition \ref{ailocdet}. 
Indeed, for some $\Rc \subset \mathfrak{P}({\O}),$ $L^0_{\Rc}(\Pi(\Phi),\Bc_c (\O))$ is the set of initial endowment $X (\cdot) \in L^0(\R,\Bc_c (\O))$ such that there exists some $\theta \in \R^d$ satisfying $X(\o) \geq  \Phi( \o,\cdot, \theta)$  $\Pc(\o) {\rm -q.s.}$ for $\o$ in a $\Rc$-full measure set (see \eqref{eqcaselec}) and we want to obtain a universally measurable version of $\theta$. 
Again the proofs are based on Theorem \ref{Essup}. 
\begin{proposition}
\label{zorro2}
Assume that Assumption \ref{anaone}  holds true. Let  $\Phi : \Omega \times   \widetilde{\O} \times \R^d \to \R \cup \{+\infty\}.$  Assume that for all $\theta \in \R^d$, $\Phi( \cdot,\cdot,\theta):(\o,\widetilde{\o}) \mapsto \Phi( \o,\widetilde{\o},\theta)$ is usa.   Recall  $\f$ from Proposition \ref{zorro1}. 
Then,  we have for all $\o \in \O$  that 
\begin{eqnarray}
\label{eqpgZ2}
\Pi(\Phi)(\o)
& =& \left\{\f(\o,\theta) : \; \theta\in \R^d \right\}+\R_+, \\
\label{eqpgZZ2}
\pi(\Phi)(\o) & =& \inf_{ \theta  \in \R^d } \f(\o,\theta). 
\end{eqnarray}
Assume that Assumption \ref{hypophi2} holds true. Then $\o \mapsto \pi(\Phi)(\o) $ is usa. \\
Assume now that Assumption \ref{hypophi3} holds true and that $\theta \mapsto \Phi( \o,\cdot,\theta)$ is $\Pc(\o)$-q.s. lsc for all $\o \in \Omega$. Let $\Rc \subset \mathfrak{P}({\O})$. Then,
\begin{eqnarray}
\label{laballebis}
	L^0_{\Rc}(\Pi(\Phi),\Bc_c (\O))  & = &  {\rm proj}_{L^0(\R,\Bc_c (\O))} L_{\Rc}^0({\rm epi}_{\Pc} \Phi,\Bc_c (\O))\\
\nonumber 
 & = & \{ X \in L^0(\R,\Bc_c (\O)): \; \exists \theta \in L^0(\R^d,\Bc_c (\O)),\; \\
 \nonumber 
  & & (\theta(\cdot),X(\cdot)) \in {\rm epi}_{\Pc} \Phi(\cdot) \; \Rc {\rm -q.s.}\}.
\end{eqnarray}
\end{proposition}
 {\sl Proof.}\\
 \noindent{\it Proof of \eqref{eqpgZ2}  and \eqref{eqpgZZ2}.} \\ 
For all $\o \in \O$, the definition of $\Pi(\Phi)(\o)$ in \eqref{laballePhi} together with \eqref{alea} show that 
\begin{eqnarray*}
\Pi(\Phi)(\o) & = & \cup_{\theta \in \R^d }  \{x\in  \R: \;    x \ge \Phi (\o,\cdot,\theta) \, \Pc(\o) {\rm -q.s.}\} =  \cup_{\theta \in \R^d }[\f(\o,\theta),\infty)
\end{eqnarray*}
and  \eqref{eqpgZ2}  and \eqref{eqpgZZ2} follow. \\
\noindent{\it If Assumption \ref{hypophi2} holds true, then $\pi(\Phi)$ is usa.} \\ 
Let $a \in \R$
\begin{eqnarray}
\nonumber
&\{\omega \in \Omega: \;  a> \pi(\Phi)(\o)  \}\\
\nonumber
=\  &\{\omega \in \Omega: \;  \ \exists (\theta,\e) \in \R^d \times  \R_+ \text{ s.t. }   a-\epsilon \geq  \Phi( \o,\cdot, \theta)   \;   \, \Pc(\o) {\rm -q.s.}\}\\
\label{cochon}
=\  &\{\omega \in \O: \;  \ \exists (\theta,\e) \in \Q^d \times  \QQ_+ \text{ s.t. }  a-\epsilon \geq \Phi( \o,\cdot, \theta)    \;   \, \Pc(\o) {\rm -q.s.}\}\\
\nonumber
=\ & \cup_{(\theta,\epsilon) \in \QQ^d \times \QQ_+} \{\omega \in \O: \; a-\epsilon \geq  \Phi( \o,\cdot, \theta) \;   \, \Pc(\o) {\rm -q.s.}\}\\
\label{vache}
=\ & \cup_{(\theta,\epsilon) \in \QQ^d \times \QQ_+}\{\omega \in \O: \   
a-\epsilon \geq  \f(\o,\theta)\}.
\end{eqnarray}
The proof of \eqref{cochon} is similar to the proof of \eqref{chat} and thus omitted.  As $\o \mapsto \f(\o,\theta)$ is usa (see Proposition \ref{zorro1}),  $\{\pi(\Phi) \geq  a \} \in \Ac(\O)$. \\
\noindent{\it If Assumption \ref{hypophi3} holds true and $\theta \mapsto \Phi( \o,\cdot,\theta)$ is $\Pc(\o)$-q.s. lsc for all $\o \in \Omega$, then \eqref{laballebis} holds true.} \\ 
Let $X \in L^0(\R^d,\Bc_c (\O))$ and let $\Gamma$ be defined as follows
\begin{eqnarray*} 
\Gamma & : = & \{(\o,\theta) \in \O \times \R^d:  \;   X(\o) \geq  \Phi( \o,\cdot, \theta)  \, \Pc(\o) {\rm -q.s.} \}.
\end{eqnarray*}
Then using Theorem \ref{Essup}, we obtain that 
\begin{eqnarray*} 
\Gamma & = & \{(\o,\theta) \in \O \times \R^d:  \; X(\o) \geq  \f (\o, \theta) \}. 
\end{eqnarray*}
Proposition \ref{zorro1} shows that $(\o, \theta) \mapsto  X(\o) - \f (\o, \theta)$ is $\Bc_c(\O) \otimes \Bc(\R^d)$-measurable and $\Gamma  \in \Bc_c(\O) \otimes \Bc(\R^d).$ 
 Thus, using the Projection theorem (see  \cite[Theorem 3.23 p75]{CV77}) and  the Auman's theorem  (see  \cite[Corollary 1]{bv}), we get that 
 ${\rm proj}_{\O}\Gamma \in \Bc_c(\Omega)$ and that  there exists $ \hat \theta(\cdot) \in  L^0(\R^d,\Bc_c (\O))$ such that    $(\o,\hat \theta(\o)) \in \Gamma$ for all $\o \in {\rm proj}_{\O}\Gamma.$ \\ 
Now, we show \eqref{laballebis}. 
Let $X\in L_{\Rc}^0(\Pi(\Phi),\Bc_c (\O))$.   Recalling  \eqref{eqcaselec}, we can fix $\o$ in the $\Rc$-full measure set where $X(\o) \in \Pi(\Phi)(\o)$. Then,  there exists $\theta \in \R^d,$ such that  $X(\o)  \ge \Phi( \o,\cdot, \theta) \, \Pc(\o){\rm -q.s.}$ and $\o \in {\rm proj}_{\O} \Gamma.$ 
 So, $(\o,\hat \theta (\o)) \in \Gamma,$ i.e.  $X(\o) \geq \Phi( \o,\cdot, \hat \theta (\o))  \, \Pc(\o) {\rm -q.s.}$  and $(\hat \theta(\cdot),X(\cdot))$ belongs to ${\rm epi}_{\Pc} \Phi(\cdot) \; \Rc {\rm -q.s.}$  
  Reversely, if $X \in {\rm proj}_{L^0(\R,\Bc_c (\O))} L_{\Rc}^0({\rm epi}_{\Pc} \Phi,\Bc_c (\O)),$ there exists  some $\theta \in L^0(\R^d,\Bc_c (\O)),$ such that $(\theta(\cdot),X(\cdot)) \in {\rm epi}_{\Pc} \Phi(\cdot) \; \Rc {\rm -q.s.}$
Fix $\o$ in the $\Rc$-full measure set where this holds true.  Then,  
$ X(\o)  \ge \Phi( \o,\cdot, \theta (\o)) \, \Pc(\o){\rm -q.s.}$ and $X(\o) \in \Pi(\Phi)(\o)$ choosing 
$\theta= \theta (\o)$. 
$\Box$\\

The next proposition gives conditions for $\Pi(\Phi)(\o)$ to be  closed and thus $\pi(\Phi)(\o) \in \Pi(\Phi)(\o)$. In the financial context, this means that the cost $\pi(\Phi)(\o)$ is in fact a price as there exists some associated super-hedging strategy. 
\begin{proposition}
\label{zorro3}
Assume that Assumption \ref{anaone} holds true. Let  $\Phi : \Omega \times   \widetilde{\O} \times \R^d \to \R \cup \{+\infty\}.$  Assume furthermore that for all $\theta \in \R^d$, $(\o,\widetilde{\o}) \mapsto \Phi( \o,\widetilde{\o},\theta)$ is usa. 
Recall  $\f$ from  Proposition \ref{zorro1}. 
Fix $\o \in \Omega$. \\
Assume that  $\theta \mapsto \Phi( \o,\cdot,\theta)$ is $\Pc(\o)$-q.s. lsc and  
that $\theta \mapsto \f(\o,\theta)$ is level coercive.  
Then,  $\Pi(\Phi)(\o)$ is closed. \\
Assume that $\theta \mapsto \Phi( \o,\cdot,\theta)$ is $\Pc(\o)$-q.s. convex and that  $\Pi(\Phi)(\o)$ is nonempty and  closed. Then,  $\lim_{|\theta| \to \infty} \frac{\f(\o,\theta)}{|\theta|}\geq0$. 
\end{proposition}
\begin{remark}
\label{remferm}
To get that $\Pi(\Phi)(\o)$ is closed, we may assume that $\theta \mapsto \Phi( \o,\cdot,\theta)$ is $\Pc(\o)$-q.s.   level coercive and lsc, 
 see Proposition \ref{zorro1}. Moreover, under the assumption that $\theta \mapsto \Phi( \o,\cdot,\theta)$ is $\Pc(\o)$-q.s. 
lsc and convex, we don\rq{}t get the equivalence between $\Pi(\Phi)(\o)$ closed and $\f(\o,\cdot)$  level coercive. Indeed, if $\Pi(\Phi)(\o)$ is closed, we only get  that  $\lim_{|\theta| \to \infty} \frac{\f(\o,\theta)}{|\theta|}\geq0.$
\end{remark}
 {\sl Proof of Proposition \ref{zorro3}.}
Assume that $\theta \mapsto \Phi( \o,\cdot,\theta)$ is $\Pc(\o)$-q.s. lsc and that  $\theta \mapsto \f(\o,\theta)$  is level coercive. 
If $\Pi(\Phi)(\o)=\emptyset$, then $\Pi(\Phi)(\o)$ is of course closed. Else, 
let $(x_n)_{n \geq 1} \subset \Pi(\Phi)(\o)$ such that $x_n$ converges to some $x^*$ when $n$ goes to infinity. We show that $x^* \in \Pi(\Phi)(\o).$
For all $n\in \N$, there exists $\theta_n\in \R^d$ such that  $ x_n\geq 
 \Phi( \o,\cdot, \theta_n) \, \Pc(\o) {\rm -q.s.}$  
 or equivalently thanks to Theorem \ref{Essup} such that  $ x_n\geq \f( \o,\theta_n).$ 
Now, if (up to a subsequence)  $|\theta_n|$ goes to infinity, dividing by $|\theta_n|$ and letting $n$ go to infinity, we get that 
$$ 0\geq \liminf_{n \to \infty} \frac{\f( \o,\theta_n)}{|\theta_n|},$$
 a  contradiction to $\f(\o,\cdot)$ level coercive. So,   there exists some $c>0$ such that $|\theta_n|\leq c$ and  (up to a subsequence) there exists some $\theta^* \in \R^d$ such that $\theta_n$ converges to $\theta^*$ when $n$ goes to $\infty.$ 
As $\theta \mapsto \Phi( \o,\cdot,\theta)$ is $\Pc(\o)$-q.s. lsc, Proposition \ref{zorro1} shows that  $\theta \to \f( \o,\theta)$ is lsc and 
 $$ x^*\geq \liminf_{n \to \infty} \f( \o,\theta_n) \geq \f( \o,\theta^*). $$ 
 So, \eqref{eqpgZ2} or Theorem \ref{Essup} show that 
$x^*\in \Pi(\Phi)(\o),$ which is thus closed. 

Assume now that $\theta \mapsto \Phi( \o,\cdot,\theta)$ is $\Pc(\o)$-q.s. convex and that  $\Pi(\Phi)(\o)$ is nonempty and closed.
Then, $\pi(\Phi)(\o)<\infty$ and there exists $\theta^* \in \R^d$ such that  $ \pi(\Phi)(\o) \geq 
\Phi( \o,\cdot, \theta^*) \, \Pc(\o) {\rm -q.s.}$
Let $\theta \in \R^d$ such that $|\theta|>1.$  Then, Theorem \ref{Essup} and the convexity property lead to 
\begin{eqnarray*}
\scalemath{0.95}
{
\left(1- \frac1{|\theta|}\right)\pi(\Phi)(\o) +\frac1{|\theta|} \f(\o,\theta) }& \geq  &\scalemath{0.95}
{ \left(1- \frac1{|\theta|}\right) \Phi( \o,\cdot, \theta^*)  + 
\frac1{|\theta|}\Phi( \o,\cdot, \theta) \;  \Pc(\o) {\rm -q.s.} }\\ 
& \geq  &  \Phi \left( \o,\cdot, \left(1- \frac1{|\theta|}\right)\theta^* +\frac1{|\theta|}\theta \right)   \;  \Pc(\o) {\rm -q.s.} \\ 
\end{eqnarray*}
It follows that $(1- \frac1{|\theta|})\pi(\Phi)(\o) +\frac1{|\theta|} \f(\o,\theta) \in \Pi(\Phi)(\o)$ and 
 $$\left(1- \frac1{|\theta|}\right)\pi(\Phi)(\o) +\frac1{|\theta|} \f(\o,\theta) \geq \pi(\Phi)(\o).$$
So, $$\liminf_{|\theta| \to \infty} \frac{\f(\o,\theta)}{|\theta|}\geq \liminf_{|\theta| \to \infty} \frac{\pi(\Phi)(\o)}{|\theta|}=0. $$
$\Box$

\section{One-period financial applications}
\label{applifi}
\subsection{Super-hedging}
\label{onegene} 
We consider a general one period financial market with $d$ risky assets and one perfectly liquid num\'eraire, which cost is set equal to one. The trading costs of the $d$ risky assets are given by real-valued cost functions  $ \Sigma=(\Sigma_0, \Sigma_1)$, where 
 $ \Sigma_0 :\O  \times 
 \R^{d}  \to  \R$  and $ \Sigma_1 :\O  \times \tilde \O \times 
 \R^{d}  \to   \R$. 
 Let $\theta^0 \in \R$ be the number of units of num\'eraire  and $\theta \in  \R^{d}$ be the number of share in the $d$ risky assets. 
 The interpretation of the cost functions is that buying a portfolio $(\theta^0, \theta)\in  \R^{d+1}$ at time $0$ and in the state $\o \in \O$  costs $\theta^0 + \Sigma_0(\o,  \theta)$ units of cash, while  buying a portfolio $(\theta^0, \theta)\in  \R^{d+1}$ at time $1$ and in the state $(\o, \tilde \o) \in \O \times \tilde{\O}$ costs $\theta^0 + \Sigma_1 (\o, \tilde \o,\theta)$ units of cash. 
This is a very general formulation which contains in particular frictionless markets, markets with transaction costs or bid-ask spreads  and also markets with illiquidity effects as the ones of  \cite{Pen11FS}. 
Let $\Rc \subset \mathfrak{P}({\O})$ and set
\begin{eqnarray*}
 \mathcal{Q}:=\{ R \otimes q:\;  R \in \Rc, q \in \mathcal{S}K,
 q(\cdot,\o) \in \Pc(\o)\;  \;  \forall \; \o \in \Omega\},
\end{eqnarray*}
where $\mathcal{S}K$ is the set of universally measurable stochastic kernels on $\tilde \O$ given $\O$ (see \cite[Definition 7.12  p134, Lemma 7.28 p174]{bs}).

We  consider a contingent claim $Z  :\O \times   \widetilde{\O} \to \R \cup \{+\infty\}$. Fix $\o \in \O$. We start with an initial cash $x$ which allows to  buy a portfolio $(\theta^0, \theta)\in  \R^{d+1}$ at time $0$, i.e. $x\geq  \theta^0 + \Sigma_0(\o,  \theta)$. 
We say that $\theta \in  \R^{d}$ is a super-hedging strategy for $Z$ if at time $1$, if we can liquidate $(\theta^0, \theta)$ and deliver $Z$ at a negative cost $\mathcal{P}(\o)$-q.s. : 
$$-\theta^0 + \Sigma_1 (\o, \cdot, -\theta) + Z(\o,\cdot) \leq 0 \;\mathcal{P}(\o) {\rm -q.s.}$$
This implies that (and is equivalent to if  $x=  \theta^0 + \Sigma_0(\o,  \theta)$) 
$$x - \Sigma_0(\o,  \theta) \geq Z(\o,\cdot)+  \Sigma_1 (\o,\cdot, -\theta)\;\mathcal{P}(\o) {\rm -q.s.}$$
So, we define the set of super-hedging prices $x$ of the contingent claim $Z$ as follows. 
\begin{definition}
\label{defpgs}
Let $Z  :\O \times   \widetilde{\O} \to  \R \cup \{+\infty\}$. 
The set-valued mapping $\Pi^{\Sigma}(Z): \Omega \twoheadrightarrow \R$ of super-hedging prices of the contingent claim $Z  $ is defined by
\begin{eqnarray*}
\scalemath{0.95}
{\Pi^{\Sigma}(Z)(\o):=\{x\in  \R: \;   \exists\, \theta\in\R^d,\;   x - \Sigma_{0} (\o, \theta)\geq Z(\o,\cdot) +  \Sigma_{1} (\o,\cdot, -\theta) \;\mathcal{P}(\o) {\rm -q.s.}\}.}
\end{eqnarray*}
The super-hedging cost of $Z$ is then defined by 
$$\pi^{\Sigma}(Z)(\o):=\inf \Pi^{\Sigma}(Z)(\o).$$
\end{definition}
Note that if $\Pi^{\Sigma}(Z)(\o)$ is not closed, then $\pi^{\Sigma}(Z)(\o)$ is not a price as it is not the initial value of some super-hedging strategy.
\begin{proposition}
\label{flash}
Assume that Assumption \ref{anaone}  holds true.  Let  $Z : \Omega \times   \widetilde{\O}  \to \R \cup \{+\infty\}$ and define $ \Phi^{\Sigma}: \Omega \times \tilde \Omega  \times \R^{d} \to\R $  as follows 
\begin{eqnarray}
\label{laballe5}
\Phi^{\Sigma}(\o,\tilde \o,\theta):= \Sigma_0(\o,   \theta) +\Sigma_1(\o,\tilde \o,  - \theta).
\end{eqnarray}
Assume that $Z$ and $\Phi^{\Sigma}( \cdot,\cdot,\theta):(\o,\widetilde{\o}) \mapsto \Phi^{\Sigma}( \o,\widetilde{\o},\theta)$ for all $\theta \in \R^d$ are usa. 
We set
$$\f^{\Sigma}(Z)(\o,\theta):= \essup[\Pc]  \left(Z( \cdot,\cdot)+ \Phi^{\Sigma}( \cdot,\cdot,\theta)\right) (\o).$$
Then,  we have for $\o \in \O$  that 
\begin{eqnarray}
\label{eqflash}
\Pi^{\Sigma}(Z)(\o)
& =&\Pi(Z + \Phi^{\Sigma})(\o) = \left\{\f^{\Sigma}(Z)(\o,\theta) : \; \theta\in \R^d \right\}+\R_+, \\
\label{eqflash1}
\pi^{\Sigma}(Z)(\o) & =& \inf_{ \theta  \in \R^d } \f^{\Sigma}(Z)(\o,\theta). 
\end{eqnarray}
Assume that Assumption \ref{hypophi2} holds true for $\Phi^{\Sigma} +Z$\footnote{This holds true if $Z$ is $\Pc(\o)$-q.s. bounded from below and  Assumption \ref{hypophi2} holds true for $\Phi^{\Sigma}$.}. Then $\o \mapsto \pi^{\Sigma}(Z)(\o) $ is usa. \\
Let $\o \in \Omega$. If  $\theta \mapsto \Phi^{\Sigma}( \o,\cdot,\theta)$ is $\Pc(\o)$-q.s. lsc and level coercive,  then $\Pi^{\Sigma}(Z)(\o)$ is closed. 
\end{proposition}
{\sl Proof.} 
The equality   $\Pi^{\Sigma}(Z)(\o)=\Pi(Z+\Phi^{\Sigma})(\o)$ follows from \eqref{laballePhi}.  
Thus, \eqref{eqflash} and \eqref{eqflash1} are direct consequences of   \eqref{eqpgZ2} and \eqref{eqpgZZ2} in Proposition \ref{zorro2}. 
Moreover, as $\Phi^{\Sigma} +Z$ satisfies Assumption \ref{hypophi2},
 Proposition \ref{zorro2} shows that $\pi^{\Sigma}(Z)$ is usa.  
 We prove the last assertion. If $\o \in \{\o \in \O : \;  Z(\o,\cdot)=\infty  \;\mathcal{P}(\o) {\rm -q.s.}\}$, $\Pi^{\Sigma}(Z)(\o) =\emptyset$ is closed. Else, 
the set $\Pi^{\Sigma}(Z)(\o)$ is closed if $\theta \mapsto Z(\o,\cdot) +\Phi^{\Sigma}( \o,\cdot,\theta)$ is $\Pc(\o)$-q.s. lsc and level coercive, see Remark \ref{remferm}. As   $\theta \mapsto \Phi^{\Sigma}( \o,\cdot,\theta)$ is $\Pc(\o)$-q.s. lsc the same holds true for $\theta \mapsto Z(\o,\cdot) +\Phi^{\Sigma}( \o,\cdot,\theta)$. Moreover, for all $(\theta_n)_{n \geq 1} \subset \R^d$ such that $|\theta_n| \to \infty$, 
$$ \liminf_{n \to \infty} \frac{\Phi^{\Sigma}(\o, \cdot, \theta_n)+ Z(\o,\cdot)}{|\theta_n|}= \liminf_{n \to \infty} \frac{\Phi^{\Sigma}( \o,\cdot, \theta_n)}{|\theta_n|} \; \;\mathcal{P}(\o) {\rm -q.s.}  $$
and we conclude as $\theta \mapsto \Phi^{\Sigma}( \o,\cdot,\theta)$ is $\Pc(\o)$-q.s. coercive. 
$\Box$\\

We study now a  situation where the cost functions depend of the initial state $\o$, of the strategy $\theta,$  of some function $y$ taken in $\o$ at time $0$ and of some function $Y$ taken in $(\o,\tilde \o)$ at time $1$. This is still a very general setup. 
This leads to the following assumption. 
\begin{assumption}
\label{borone} Let $y:\O  \to \R^d$   and  $Y:\O \times   \widetilde{\O} \to \R^d$ be  Borel measurable functions. Let 
$S_0, \, S_1:  \Omega \times  \R^d  \times  \R^d \to \R$ and $g: \Omega \times \R^d\to \R \cup \{+\infty\}.$ \\
We assume that $g$ is usa   and that $(\o,z) \mapsto S_1(\o,z,\theta)$ is usa for all $\theta \in \R^d$. We set 
\begin{eqnarray*}
\Sigma_0(\o,\theta) &:=& S_0(\o,y(\o), \theta)\\
\Sigma_1(\o,\tilde \o, \theta) &:=& S_1(\o,Y(\o,\tilde \o), \theta)\\
Z(\o,\tilde \o) & := & g(\o,Y(\o,\tilde \o)).
\end{eqnarray*}
\end{assumption}
In this context, we are able to express the super-hedging prices and the super-heding cost as dual and bi-dual of the payoff $g$ relative to the support  of $Y$, ${\rm supp}_{\Pc}Y$. 
\begin{proposition}
\label{black}
Assume that Assumptions \ref{anaone} and \ref{borone} hold true.  
Fix some  $\o \in \O$  and suppose that $g(\o,\cdot)$ and $S_1(\o,\cdot,\theta)$  for all  $\theta \in \R^d$ are lsc. Then,  we get that
\begin{eqnarray}
\label{eqf*}
\f^{\Sigma}(Z)(\o,\theta) & = & f^{S_1}(\o,-\theta ) +S_0(\o,y(\o), \theta),
 \end{eqnarray}
where $f$ and $f^{S_1}$ are defined  by
\begin{eqnarray*}
f(\o,z)& := &-g(\o,z)+\delta_{{\rm supp}_{\Pc}Y(\o)}(z)\\
f^{S_1}(\o,x) & := & \sup_{z\in \R^d}\left(S_1(\o,z,x)-f(\o,z)\right),
\end{eqnarray*}
and $\delta_{C(\o)}(z)=0,$ if $z\in C(\o)$ and $+\infty$ else. \\
We have also that 
 \begin{eqnarray*}
\nonumber
\pi^{\Sigma}(Z)(\o)& =& -(f^{S_1})^{S_0}(\o,y(\o)),
\end{eqnarray*}
where for all $h:\O \times \R^d \to \R \cup \{+\infty\}$, $h^{S_0}$ is defined by 
\begin{eqnarray*}
h^{S_0}(\o,z) & := & 
\sup_{ x  \in \R^d }   \left( -S_0(\o,z, -x) -h(\o,x)\right).
\end{eqnarray*}
\end{proposition}
{\sl Proof.}
Let $x\in \Pi^{\Sigma}(Z)(\o).$ There exists $ \theta\in\R^d,$ such that 
$$x - S_0(\o,y(\o), \theta)\geq g(\o,Y(\o,\cdot)) +  S_1(\o,Y(\o,\cdot), -\theta) \;\mathcal{P}(\o) {\rm -q.s.}$$
As $Y$ is Borel measurable and $g$ and $(\o,z) \mapsto S_1(\o,z,\theta)$ are usa,  $(\o,\tilde \o) \mapsto g(\o,Y(\o,\tilde \o))$ and $(\o,\tilde \o) \mapsto S_1(\o,Y(\o,\tilde \o), -\theta)$  are 
usa (see \cite[Lemma 7.30 p177]{bs}). So, Theorem \ref{Essup} implies that 
$$x - S_0(\o,y(\o), \theta)\geq \essup[\Pc]  \left[g(\cdot,Y(\cdot)) +  S_1(\cdot,Y(\cdot), -\theta)\right] (\o).$$
As $g(\o,\cdot)$  and $S_1(\o,\cdot,-\theta)$ are lsc,    Proposition \ref{lemma-essup-h(X)} implies that 
\begin{eqnarray*}
\essup[\Pc]  \left[g(\cdot,Y(\cdot)) +  S_1(\cdot,Y(\cdot), -\theta)\right]) (\o)
&= & \sup_{z\in  {\rm supp}_{\Pc}Y(\o)}\left(g(\o,z) + S_1(\o,z, -\theta)\right) \\
&= & \sup_{z\in \R^d}\left(-f(\o,z) + S_1(\o,z, -\theta)\right)\\
 &= & f^{S_1}(\o,-\theta). 
\end{eqnarray*}
So, we conclude that \eqref{eqf*} holds true. 
Moreover, using \eqref{eqflash1}
\begin{eqnarray*}
\pi^{\Sigma}(Z)(\o) 
 & =& \inf_{ \theta  \in \R^d }   \left(f^{S_1}(\o,-\theta) +S_0(\o,y(\o), \theta)\right)\\
  & =& -\sup_{ x \in \R^d }   \left(-f^{S_1}(\o,x) -S_0(\o,y(\o), -x)\right)\\
  & =  &-(f^{S_1})^{S_0}(\o,y(\o)).
\end{eqnarray*}
$\Box$\\


We want to find conditions guaranteeing that $f^{S_1}$ is the  Fenchel-Legen-\\dre conjugate of $f$ and $(f^{S_1})^{S_0}$ is the Fenchel-Legendre biconjugate of $f$ and where the Fenchel-Legendre biduality theorem applies. 
First, we need to assume that $S_0=S_1=:S,$  $S(\o,z, -x)=-S(\o,z, x)$ and $S(\o,x, z)=S(\o,z, x).$ Then, it follows that 
\begin{eqnarray*}
f^{S_1}(\o,x) & = & \sup_{z\in \R^d}\left(S(\o,z,x)-f(\o,z)\right)\\
h^{S_0}(\o,z) & = & 
\sup_{ x  \in \R^d }   \left( S(\o,x, z) -h(\o,x)\right).
\end{eqnarray*}
The Fenchel-Legendre biduality  theorem is based on the fact that a proper lsc convex function $h$ is  the pointwise supremum of its affine supports, i.e. of the affine functions that are lower or equal than $h$ everywhere and equal at some point, see \cite[Theorem 8.13 p309]{rw}.  Thus, we need to postulate that $S(\o,z, x)$ is an affine function, i.e. 
$S(\o,z, x)=(D(\o)x)z$ for some $d \times d$-matrix $D(\o)$. 
Recalling that $S$ is symmetric, $D(\o)$ needs to be a diagonal matrix. 
It follows that 
\begin{eqnarray*}
\Sigma_0(\o,\theta) & =  & (D(\o)\theta)y(\o) =\theta (D(\o)y(\o))\\ 
 \Sigma_1(\o,\tilde \o, \theta) & = &  (D(\o)\theta)Y(\o,\tilde \o) = \theta (D(\o)Y(\o,\tilde \o)). 
\end{eqnarray*}
So, changing $y(\o)$ into $D(\o)y(\o)$ and $Y(\o,\tilde \o)$ into $D(\o)Y(\o,\tilde \o)$ 
we are back to a frictionless market where the market price of the assets at time $0$ (resp. time $1$) is $D(\o)y(\o)$ (resp. $D(\o)Y(\o,\tilde \o)$). This is what we do in the next assumption. 
\begin{assumption}
\label{bortwo} Let $y:\O  \to \R^d$ and  $Y:\O \times   \widetilde{\O} \to \R^d$ be  Borel measurable functions. We set 
\begin{eqnarray*}
S(\o,x, z) &  := & zx\\
\Sigma_0(\o,\theta) &:=& S(\o,y(\o), \theta)=\theta y(\o)\\
\Sigma_1(\o,\tilde \o, \theta) &:=& S(\o,Y(\o,\tilde \o), \theta)=\theta Y(\o,\tilde \o).
\end{eqnarray*}
\end{assumption}
Recall that the (upper) closure $\overline{h}$ of $h$ is the smallest upper-semicontinuous (usc)   function which dominates $h,$ i.e. 
$\overline{h}(x)=\limsup_{y \to x} h (y)$.  The lower closure $\underline{h}$ of $h$ is defined symmetrically, see  \cite[1(7) p14]{rw}.
\begin{corollary}
\label{lempart1}
Assume that Assumptions \ref{anaone} and \ref{bortwo} hold true. Let  $g: \Omega \times \R^d\to \R \cup \{+\infty\}$ be  usa   and  set $Z(\o,\tilde \o) :=  g(\o,Y(\o,\tilde \o)).$   
Fix some  $\o \in \O.$  Suppose that $g(\o,\cdot)$ is lsc and recall $f$ from Proposition \ref{black}. Let $f^*$ be its Fenchel-Legendre conjugate  and $f^{**}$ be its Fenchel-Legendre biconjugate: 
\begin{eqnarray*}
f^*(\o,x) & :  = & \sup_{z\in \R^d}\left(xz -f(\o,z)\right)\\
 f^{**}(\o,x) &  = & \sup_{z\in \R^d}\left(xz -f^*(\o,z)\right).
\end{eqnarray*}
Then,  we get that
for all $\theta \in \R^d$
\begin{eqnarray*}
\f^{\Sigma}(Z)(\o,\theta) & = & f^*(\o,-\theta ) +\theta y(\o)\\
\pi^{\Sigma}(Z)(\o)& =& -f^{**}(\o,y(\o)). 
 \end{eqnarray*}
Moreover, suppose that $g(\o,\cdot)$ is proper\footnote{The effective domain of $g(\o,\cdot)$ is defined by
${\rm dom \,} g(\o,\cdot):=\{x \in \R^d: \, g(\o,x)< \infty\}$ and $g(\o,\cdot)$ is proper if dom $g(\o,\cdot) \neq \emptyset$ and $g(\o,x)>- \infty$  for all $x \in \R^d$.} and that there exists some concave function $\varphi$  such that $g(\o,\cdot)\le \varphi<\infty$ on ${\rm conv}\,{\rm supp}_{\Pc}Y(\o),$
where ${\rm conv}\,{\rm supp}_{\Pc}Y(\o)$ is the convex envelope of ${\rm supp}_{\Pc}Y(\o),$ i.e. the smallest convex set that contains ${\rm supp}_{\Pc}Y(\o)$.
We have that 
 \begin{eqnarray}
\label{leprix}
\pi^{\Sigma}(Z)(\o)
&= &    \overline{{\rm conc}}(g,{\rm supp}_{\Pc}Y)(\o,y(\o))-\delta_{ {\rm conv}\,{\rm supp}_{\Pc}Y(\o)}(y(\o)),
\end{eqnarray}
where the relative concave envelope of $g(\o,\cdot)$ with respect to ${\rm supp}_{\Pc}Y(\o)$ is given by 
 \begin{eqnarray*}
\scalemath{0.95}{ 
{\rm conc}(g, {\rm supp}_{\Pc}Y)(\o,x) := 
\inf\{ v(x): \;  v\,{\rm\, is\, concave\, and\, }v(z)\ge g(\o,z),\, \forall z\in {\rm supp}_{\Pc}Y (\o)\}.
}
\end{eqnarray*} 
\end{corollary}
{\sl Proof.}
Recall that $\o \in \O$ is fixed. The first part of the corollary is a direct application of Proposition \ref{black}. 
Under the additional assumptions that $g(\o,\cdot)$ is proper and that there exists some concave function $\varphi$  such that $g(\o,\cdot)\le \varphi<\infty$ on ${\rm conv}\,{\rm supp}_{\Pc}Y(\o)$,  ${\rm conv \,}f(\o,\cdot)$ satisfies the following equality (this is proved as in \cite[(2.11) Theorem 2.8]{BCL}):
\begin{eqnarray*}
{\rm conv \,}f (\o,\cdot) &=&-{\rm conc}(g, {\rm supp}_{\Pc}Y)(\o,\cdot) +\delta_{{\rm conv}\,{\rm supp}_{\Pc}Y(\o)}(\cdot).\end{eqnarray*}
As   ${\rm conv}\,{\rm supp}_{\Pc}Y(\o)$ is nonempty,  ${\rm conv \,}f(\o,\cdot)$ is  proper and
\cite[Theorem 11.1 p474]{rw} implies that  $f^*(\o,\cdot)$ is proper, lsc and convex and that the biduality relation holds true:
$
f^{**}(\o,z)= \underline{{\rm conv}} \,f(\o,z).$
Finally, we obtain that 
 \begin{eqnarray*}
\pi^{\Sigma}(Z)(\o)& =&-f^{**}(\o,y(\o))= -\underline{{\rm conv}} \,f(\o,y(\o)).
\end{eqnarray*}
$\Box$\\
We now prove that if $y$ and $Y$ are non-negative,  Assumption \ref{hypophi3} holds true for $\Phi^{\Sigma}$ in the frictionless case. 
\begin{lemma}
\label{unifcont}
Assume that Assumption \ref{anaone} holds true. Let $y:\O  \to \R^d_+$ and    $Y:\O \times   \widetilde{\O} \to \R^d_+. $ Assume that  
\begin{eqnarray*}
\Sigma_0(\o,\theta) &:=& \theta y(\o)\\
\Sigma_1(\o,\tilde \o, \theta) &:=& \theta Y(\o,\tilde \o).
\end{eqnarray*}
Then,  Assumption \ref{hypophi3} holds true for $\Phi^{\Sigma}$. 
\end{lemma}
{\sl Proof.}
Here $\Phi^{\Sigma}( \o,\widetilde{\o},\theta)=  -\theta  \Delta Y(\o,\widetilde{\o})$ where $ \Delta Y(\o,\widetilde{\o}):= Y(\o,\widetilde{\o}) -y(\o).$ \\
Fix $\o \in \O$. 
If $\max_{i \in \{1,\dots, d\}}y_i(\omega)=0,$  then  $ \theta y(\o) =0$ for all $ \theta \in \R^d$. Let $\eta=1$. 
Fix ${\epsilon}>0$ and $\theta \in \R^d$. 
For all $\tilde \theta  \in [0, \infty)^d,$ $\Pc(\omega)$-q.s.  we obtain that 
\begin{align*}
\Phi^{\Sigma}( \o,\cdot,\tilde \theta + \theta) - \Phi^{\Sigma}( \o,\cdot, \theta)=- \tilde \theta \Delta Y(\o,\cdot)= - \tilde \theta Y(\o,\cdot) \leq 0 \leq \e. 
\end{align*}
since $Y_i \geq 0.$ 
We can choose $\tilde{\theta}$ such that $\theta+\tilde{\theta} \in \QQ^d$ and $|\tilde \theta| \leq \e \eta=\e$ and Assumption \ref{hypophi3} holds true. \\
Else, $\max_{i \in \{1,\dots, d\}}y_i(\omega)>0$ and let $\eta = \frac{1}{\max_{i \in \{1,\dots, d\}}y_i(\omega)}>0$. 
Fix ${\epsilon}>0$ and $\theta \in \R^d$. 
Choose  $\tilde{\theta} \in [0, \infty)^d$ such that 
\begin{align*}
\tilde{\theta}^1+ \dots +\tilde{\theta}^d \le \epsilon \eta= \frac{\epsilon}{\max_{i \in \{1,\dots, d\}}y_i(\omega)}.
\end{align*}
Then, as $\tilde \theta  \in  [0, \infty)^d,$ $|\tilde \theta| \leq \e \eta.$  Moreover, $\tilde \theta y(\o) \le \epsilon$. As $Y_i \geq 0$, it follows that  $\Pc(\omega)$-q.s. 
\begin{align*}
\Phi^{\Sigma}( \o,\cdot,\tilde \theta + \theta) - \Phi^{\Sigma}( \o,\cdot, \theta)= - \tilde \theta Y(\o,\cdot) + \tilde \theta y(\o) \leq \e. 
\end{align*}
Again, the above inequality is valid for some $\tilde{\theta}$ such that $\theta+\tilde{\theta} \in \QQ^d$ and Assumption \ref{hypophi3} holds true. $\Box$\\
\begin{lemma}
\label{unifcont1}
Assume that Assumptions \ref{anaone} and \ref{bortwo} hold true. 
Let  $Z:\O \times   \widetilde{\O} \to \R \cup \{+\infty\}$ be usa and set 
$\f^{\Sigma}(Z)(\o,\theta)$ $
:=$ $\essup[\Pc]\left(Z-\theta \Delta Y \right)(\o).$ 
Let  $\sigma_{D}(z):=\sup_{x \in D} (-xz)$ be the support function of $-D$ for any $D\subset \mathbb{R}^d$. 
Fix $\o \in \O$. \\
If $\sigma_{{\rm supp}_{\Pc}\Delta Y(\o)}(\theta)>0$ for all $\theta \in \R^d \setminus\{0\},$ then $\theta \mapsto$ $\f^{\Sigma}(Z)(\o,\theta)$  is level coercive and $\Pi^{\Sigma}(Z)(\o)$ is closed. \\
Assume now that $\sigma_{{\rm supp}_{\Pc} \Delta Y (\o)} \leq 0.$  Then,  $\Pi^{\Sigma}(Z)(\o)$ is closed and $\pi^{\Sigma}(Z)(\o) =\essup[\Pc] Z(\o)$. \\
Conversely, if $\Pi^{\Sigma}(Z)(\o)$ is nonempty and closed then, $\sigma_{{\rm supp}_{\Pc} \Delta Y (\o)}  \geq 0$. 
\end{lemma}
{\sl Proof.}
Here $\Phi^{\Sigma}( \o,\widetilde{\o},\theta)=  -\theta  \Delta Y(\o,\widetilde{\o})$ where $ \Delta Y(\o,\widetilde{\o}):= Y(\o,\widetilde{\o}) -y(\o).$ \\
So, for all $\theta \in \R^d$, $(\o,\widetilde{\o}) \mapsto Z( \o,\widetilde{\o})+ \Phi^{\Sigma}( \o,\widetilde{\o},\theta)=  Z( \o,\widetilde{\o})-\theta  \Delta Y(\o,\widetilde{\o})$ is  usa. 
The continuity and the convexity of  $\Phi^{\Sigma}( \o,\widetilde{\o},\cdot)$ is clear. 
If $\o \in \{\o \in \O : \;  Z(\o,\cdot)=\infty  \;\mathcal{P}(\o) {\rm -q.s.}\}$, then $\Pi^{\Sigma}(Z)(\o) =\emptyset$ is closed. Else, 
the set $\Pi^{\Sigma}(Z)(\o)$ is closed if $\theta \mapsto \f^{\Sigma}(Z)(\o,\theta)$ 
is level coercive, see  Proposition \ref{zorro3}. Theorem \ref{Essup} proves that 
\begin{eqnarray*}
 \f^{\Sigma}(Z)(\o,\theta) : & = & \essup[\Pc]\left(Z-\theta \Delta Y \right)(\o)  \geq  Z(\o, \cdot)-\theta \Delta Y(\o, \cdot) \;\mathcal{P}(\o) {\rm -q.s.}
 \end{eqnarray*}
 So, we obtain for all $|\theta_n| \to \infty$ that 
 \begin{eqnarray*}
\liminf_{n \to \infty} \frac{\f^{\Sigma}(Z)(\o,\theta_n)}{|\theta_n|} & \geq & \liminf_{n \to \infty} \frac{  Z(\o, \cdot)-\theta_n \Delta Y(\o, \cdot)}{|\theta_n|}=-\theta^*\Delta Y(\o, \cdot) \;\mathcal{P}(\o) {\rm -q.s.}
\end{eqnarray*}
where $|\theta^*|=1$. So,  Theorem \ref{Essup} again proves that 
\begin{eqnarray}
\label{missi}
\liminf_{n\to \infty} \frac{\f^{\Sigma}(Z)(\o,\theta_n)}{|\theta_n|} & \geq & \essup[\Pc]\left(-\theta^* \Delta Y \right)(\o).
\end{eqnarray}
Now, Proposition \ref{lemma-essup-h(X)} shows that 
\begin{eqnarray}
 \label{enfin?}
 \essup[\Pc]\left(-\theta \Delta Y \right)(\o)      =   \sup_{x \in  {\rm supp}_{\Pc}\Delta Y(\o)}\left(-\theta x \right) =  \sigma_{{\rm supp}_{\Pc}\Delta Y(\o)}(\theta).
\end{eqnarray}
So, if $\sigma_{{\rm supp}_{\Pc}\Delta Y(\o)}(\theta)>0$ for all $\theta \in \R^d \setminus\{0\},$  then \eqref{missi} shows that $\theta \mapsto \f^{\Sigma}(Z)(\o,\theta)$ 
is level coercive and $\Pi^{\Sigma}(Z)(\o)$ is closed.

Assume now that $\sigma_{{\rm supp}_{\Pc} \Delta Y (\o)} \leq 0.$  
We prove that
\begin{eqnarray}
 \label{enfin??}
\Pi^{\Sigma}(Z)(\o)=  [\essup[\Pc] Z(\o),+\infty).
\end{eqnarray}
This will imply that $\Pi^{\Sigma}(Z)(\o)$   is closed and that $\pi^{\Sigma}(Z)(\o) =\essup[\Pc] Z(\o)$. \\
First, we show that for all $\theta \in \R^d,$ $\theta \Delta Y(\o,\cdot) = 0 \, \Pc(\o) {\rm -q.s.}$ 
Fix $\theta \in \R^d.$ Using \eqref{enfin?} and Theorem \ref{Essup}, we get  that 
$$0\geq \sigma_{{\rm supp}_{\Pc}\Delta Y(\o)}(\theta) =\essup[\Pc]\left(-\theta \Delta Y \right)(\o) \geq - \theta \Delta Y(\o,\cdot)  \, \Pc(\o) {\rm -q.s.}$$
Thus, $\theta \Delta Y(\o,\cdot) \ge 0 \, \Pc(\o) {\rm -q.s.}$  Applying to $-\theta,$  $\theta \Delta Y(\o,\cdot) = 0 \, \Pc(\o) {\rm -q.s.}$ holds true. 
Let $x \in \Pi^{\Sigma}(Z)(\o)$. Then, there exists $\theta \in \R^d$ such that  $ x+\theta \Delta Y(\o,\cdot) \ge Z(\o,\cdot) \, \Pc(\o) {\rm -q.s.}$ and thus 
$ x \ge Z(\o,\cdot) \, \Pc(\o) {\rm -q.s.}$  Theorem \ref{Essup} shows  that  $ x \geq \essup[\Pc] Z (\o)$ and the first inclusion in \eqref{enfin??} is proved. 
Now let $x \geq \essup[\Pc] Z(\o)$. Theorem \ref{Essup} again implies that $ x \ge Z(\o,\cdot) \, \Pc(\o) {\rm -q.s.}$ and $x \in \Pi^{\Sigma}(Z)(\o)$ choosing $\theta=0$. This concludes the proof of 
\eqref{enfin??}.

Assume now that $\Pi^{\Sigma}(Z)(\o)$ is nonempty and closed.  Then, $ \pi^{\Sigma}(Z)(\o) \in \Pi^{\Sigma}(Z)(\o)$ and  there exists $\theta^* \in \R^d$ such that  $$ \pi^{\Sigma}(Z)(\o)+\theta^* \Delta Y(\o,\cdot) \ge Z(\o,\cdot) \, \Pc(\o) {\rm -q.s.}$$
Let $\theta \in \R^d$. Then, using \eqref{enfin?} and Theorem \ref{Essup}, we get  that 
$$\sigma_{{\rm supp}_{\Pc}\Delta Y(\o)}(\theta)=\essup[\Pc]\left(-\theta \Delta Y\right)(\o) \geq -\theta \Delta Y(\o,\cdot) \, \Pc(\o) {\rm -q.s.}$$
Thus, $$ \pi^{\Sigma}(Z)(\o) + \sigma_{{\rm supp}_{\Pc}\Delta Y(\o)}(\theta)+(\theta^* +\theta)\Delta Y(\o,\cdot) \ge Z(\o,\cdot) \, \Pc(\o) {\rm -q.s.}$$
and $ \pi^{\Sigma}(Z)(\o) + \sigma_{{\rm supp}_{\Pc}\Delta Y(\o)}(\theta) \in  \Pi^{\Sigma}(Z)(\o).$ This implies that 
$ \pi^{\Sigma}(Z)(\o) + \sigma_{{\rm supp}_{\Pc}\Delta Y(\o)}(\theta) \geq \pi^{\Sigma}(Z)(\o),$ i.e. $\sigma_{{\rm supp}_{\Pc}\Delta Y(\o)}(\theta) \geq 0$. 
 $\Box$\\

\subsection{Absence of Instantaneous Profit}

We now focus on the  notion of Absence of Instantaneous Profit (AIP),  which has been introduced in \cite{BCL} in the uni-prior case for a frictionless financial model. 
Fix $\o \in \Omega$. 
We say that there is an Instantaneous Profit (IP)  if one is quasi-sure to  make a strictly positive profit, i.e. if there exist $e_{\o}>0$  and $\theta_{\o}\in \R^d$ such that $
\Sigma_{0} (\o, \theta_{\o}) +  \Sigma_{1} (\o,\cdot, -\theta_{\o}) \le -e_{\o} \, \Pc(\o) {\rm -q.s.}$ or equivalently that if there exists $e_{\o}>0$  such that $-e_{\o} \in \Pi^{\Sigma}(0)(\o).$ 
This is also equivalent to assume that $\pi^{\Sigma}(0)(\o)<0.$ 
 \begin{definition}
 \label{defipone}
Fix $\o \in \O$.  
The Absence of Instantaneous Profit condition, denoted by AIP$(\Pc(\o)),$  holds true if and only if $\Pi^{\Sigma}(0)(\o) \cap (-\infty,0)=\emptyset$ or equivalently  if and only if $\pi^{\Sigma}(0)(\o) \geq 0.$ 
\end{definition}
\begin{remark}
The AIP$(\Pc(\o))$ condition is also equivalent to $\Pi^{\Sigma}(0)(\o) \subset \R_+.$\\
Now, assume that $\Sigma_0 (\o,0)+\Sigma_1 (\o,\cdot,0)\leq0 \; \Pc(\o)$-q.s. for $\o\in \O_z$ for some  $\O_z \subset \O$. 
Then, 
for   $\o \in \O_z$, the AIP$(\Pc(\o))$ condition is equivalent to  $\pi^{\Sigma}(0)(\o)=0$ or $\Pi^{\Sigma}(0)(\o) \cap \R_-=\{0\}.$ Moreover, 
if $\o \in \O_z$ and AIP$(\Pc(\o))$ holds true, then $\pi^{\Sigma}(0)(\o)=0 \in \Pi^{\Sigma}(0)(\o)$. Thus, $\Pi^{\Sigma}(0)(\o)$ is closed and $\Pi^{\Sigma}(0)(\o)=[0,\infty).$ 
\end{remark}
Proposition \ref{black} implies directly the following characterization of AIP$(\Pc(\o))$ in the context of Assumption \ref{borone}. 
\begin{corollary}
\label{blackbird}
Assume that Assumptions \ref{anaone} and \ref{borone} hold true.  
Fix some  $\o \in \O$  and suppose that $S_1(\o,\cdot,\theta)$ is lsc for all  $\theta \in \R^d$. Let 
\begin{eqnarray}
\label{eqfcoro}
\sigma_{{\rm supp}_{\Pc}Y(\o)} ^{S_1}(\o,\theta) & : = &  \sup_{z\in {\rm supp}_{\Pc}Y(\o)} S_1(\o,z,-\theta ).
\end{eqnarray}
Then, we get that 
\begin{eqnarray*}
\pi^{\Sigma}(0)(\o)& =& \inf_{ \theta  \in \R^d }   \left( S_0(\o,y(\o), \theta) +\sigma_{{\rm supp}_{\Pc}Y(\o)}^{S_1}(\o,\theta)\right).
\end{eqnarray*}
AIP$(\Pc(\o))$ is satisfied  if and only if for all $ \theta \in \R^d$, 
\begin{eqnarray}
\label{eqfcoro2} S_0(\o,y(\o), \theta)+ \sigma_{{\rm supp}_{\Pc}Y(\o)}^{S_1}(\o,\theta) \geq 0. 
\end{eqnarray}
So, AIP$(\Pc(\o))$ is satisfied  if and only if for all $  \theta  \in \R^d$, there exists $z_{ \theta}\in {\rm supp}_{\Pc}Y(\o)$ such that 
$$S_0(\o,y(\o),  \theta) + S_1(\o,z_{ \theta},- \theta) \geq 0. $$
\end{corollary}
Now, we provide a characterization  of AIP$(\Pc(\o))$ in the context of Assumption \ref{bortwo}. 
\begin{corollary}
\label{NGDtwo}
Assume that Assumptions \ref{anaone} and \ref{bortwo} hold true. Fix $\o \in \O$. 
Then,  the following conditions are equivalent:
\begin{enumerate}
\item AIP$(\Pc(\o))$ is satisfied.
\item  $y(\o) \in {\rm conv}\,{\rm supp}_{\Pc}Y(\o)$  or $0 \in {\rm conv}\,{\rm supp}_{\Pc}\Delta Y(\o)$.
\item $\theta y(\o) + \sigma_{{\rm supp}_{\Pc}Y(\o)} (\theta) \geq 0$ for all $\theta \in \R^d$ or  $\sigma_{{\rm supp}_{\Pc}\Delta Y(\o)} \geq 0.$     
\end{enumerate}
Here, $\sigma_{D}(z)=\sup_{x \in D} (-xz)$ is the support function of $-D.$
\end{corollary}

\begin{remark}
\label{foi}
In the case $d=1$, \cite[Lemma 2.6]{BCL} implies that the previous conditions are equivalent to
$y(\o) \in  [\essinf[\Pc] Y(\o), \essup[\Pc]  Y(\o)]\cap \R.$  
\end{remark}

{\sl Proof of Corollary \ref{NGDtwo}.}
The assumptions of Corollary \ref{lempart1} are satisfied for $g=0$
and we get that
$$\pi^{\Sigma}(0)(\o)=-\delta_{{\rm conv}\,{\rm supp}_{\Pc}Y(\o)}(y(\o)).$$ 
The AIP$(\Pc(\o))$ condition  holds true if and only if $\pi^{\Sigma}(0)(\o)\geq 0.$ \\
Hence, AIP$(\Pc(\o))$ holds true if and only if $y(\o)\in  {\rm conv}\,{\rm supp}_{\Pc}Y(\o)$ or equivalently if $0 \in {\rm conv}\,{\rm supp}_{\Pc}\Delta Y(\o)$ and 1. is equivalent to 2.\\
As Assumption \ref{bortwo} holds true, \eqref{eqfcoro} shows that $\sigma_{{\rm supp}_{\Pc}Y(\o)} ^{S_1}(\o,\theta) = \sigma_{{\rm supp}_{\Pc}Y(\o)} (\theta)$ and the equivalence between 
 1. and 3. follows for \eqref{eqfcoro2}.
$\Box$\\

We now make the link with the no-arbitrage condition. The AIP$(\Pc(\o))$ is the minimal condition in order to do pricing. Indeed, we have proved in \eqref{leprix} of Corollary \ref{lempart1} 
that if AIP$(\Pc(\o))$ fails, i.e. if $y(\o) \notin {\rm conv}\,{\rm supp}_{\Pc}Y(\o)$, then 
$\pi^{\Sigma}(Z)(\o)=-\infty$ for the contingent claim $Z=g(Y)$. Nevertheless AIP$(\Pc(\o))$   is not sufficient to solve the problem of expected utility maximization, where NA$(\Pc(\o))$ is required to obtain a well-posed problem. We show in Lemma \ref{impli} that NA$(\Pc(\o))$  implies AIP$(\Pc(\o))$  but that the reverse does not hold true. We also give conditions for the equivalence in the case of frictionless markets, see Proposition \ref{oalocal}. 
 \begin{definition}
 \label{defaoaone}
Fix $\o \in \O$. 
The no-arbitrage condition NA$(\Pc(\o))$  at $\o$ asserts that if $\Sigma_{0} (\o, \theta) +  \Sigma_{1} (\o,\cdot, -\theta) \le 0\, \Pc(\o) {\rm -q.s.}$ for some $\theta \in \R^d$, then 
$\Sigma_{0} (\o, \theta) +  \Sigma_{1} (\o,\cdot, -\theta)= 0\, \Pc(\o) {\rm -q.s.}$
\end{definition}
Thus, $\theta$ is an arbitrage at $\o$ if $\Sigma_{0} (\o, \theta) +  \Sigma_{1} (\o,\cdot, -\theta) \le 0\, \Pc(\o) {\rm -q.s.}$  and if there exists some $P\in \Pc(\o)$ such that 
$P(\Sigma_{0} (\o, \theta) +  \Sigma_{1} (\o,\cdot, -\theta)<0)>0.$ 
This result is the generalization of \cite[Lemma 2.16]{BCL} in a multiple-prior market with frictions. 
\begin{lemma}
\label{impli}
Fix $\o \in \O$. NA$(\Pc(\o))$  implies AIP$(\Pc(\o))$  but the reverse does not hold true. 
\end{lemma}
{\sl Proof.}
Assume that there is an Instantaneous Profit (IP)  at $\o$. There exist $e_{\o}>0$  and $\theta_{\o}\in \R^d$ such that $
\Sigma_{0} (\o, \theta_{\o}) +  \Sigma_{1} (\o,\cdot, -\theta_{\o}) \le -e_{\o} \, \Pc(\o) {\rm -q.s.}$  Thus, $\theta_{\o}$ is an arbitrage at $\o$. \\
On the contrary AIP$(\Pc(\o))$ may be  satisfied while NA$(\Pc(\o))$ fails. Take $d=1,$ $\Pc(\o)=\{q(\cdot,\o)\}$ for some $q(\cdot,\o) \in \mathfrak{P}(\widetilde{\O})$ such that  $\o \mapsto q(\cdot,\o) $ is Borel measurable. Using \cite[Corollary 7.14. 1 p121]{BC19}, ${\rm gph} \,\Pc={\rm gph} \,q$ is Borel and Assumption \ref{anaone} holds true. 
In the context of Assumption \ref{bortwo}, choose $y(\o)=0$ and $Y(\o, \cdot)$ that follows a uniform distribution on $[0,1]$ under $q(\cdot,\o).$
Then, 
$\Sigma_{0} (\o, \theta)=0,$  $\Sigma_{1} (\o, \tilde\o, \theta)=\theta Y(\o, \tilde\o)$ and 
$$q(\Sigma_{0} (\o, 1) +  \Sigma_{1} (\o,\cdot, -1) <0 ,\o) =q(Y(\o, \cdot)>0 ,\o)>0,$$
and there is an arbitrage at $\o$. Nevertheless, $ y(\o)=0 \in {\rm conv}\,{\rm supp}_{\Pc}\Delta Y(\o)=[0,1]$ and Corollary \ref{NGDtwo} asserts that  AIP$(\Pc(\o))$ holds true. 
$\Box$\\

The following proposition is a direct consequence of Corollary \ref{NGDtwo} and \cite[Proposition 5.7]{BC19}. 
Of course it implies Lemma \ref{impli} but necessitates  Assumptions \ref{anaone} and \ref{bortwo}. 
Recall that for a convex set $C \subset \mathbb{R}^{d}$, the relative interior of $C$ (see \cite[p64]{rw}) is ${\rm ri} \,C=\{y \in C: \, \exists \, \varepsilon >0,\;  {\rm aff}\, C \cap B(y,\varepsilon) \subset C\}$
where $B(y,\varepsilon)$ is the open ball in $\mathbb{R}^{d}$ of center $y$ and radius $\varepsilon$ and ${\rm aff}\, C$ is the affine hull of $C$, i.e. the smallest affine set that includes $C$. 
\begin{proposition}
\label{oalocal}
Assume that Assumptions \ref{anaone} and \ref{bortwo} hold true. Fix $\o \in \O$.    
\begin{enumerate}
\item The AIP$(\Pc(\o))$  condition holds true if and only if $0 \in {\rm conv}\,{\rm supp}_{\Pc}\Delta Y(\o).$
\item The NA$(\Pc(\o))$ condition holds true if and only if $0 \in{\rm ri}\,{\rm conv}\,{\rm supp}_{\Pc}\Delta Y(\o)$. 
\end{enumerate}
So, both notions coincide if  
 $0 \notin {\rm conv}\,{\rm supp}_{\Pc}\Delta Y(\o) \setminus{\rm ri}\,{\rm conv}\,{\rm supp}_{\Pc}\Delta Y(\o)$. 
\end{proposition}
Note that when $d=1,$ we can directly  show that if AIP$(\Pc(\o))$ holds true and if $y(\o) \notin \{\essinf[\Pc] Y(\o),\essup[\Pc]  Y(\o)\}$, then  NA$(\Pc(\o))$ holds true. \\
Indeed,  AIP$(\Pc(\o))$ implies that  $y(\o) \in  [\essinf[\Pc] Y(\o), \essup[\Pc]  Y(\o)] \cap \R,$  see  Remark \ref{foi}. 
Let $\theta \in \R$ be such that $\theta \Delta Y(\o,\cdot) \geq 0$ $\Pc(\o)$-q.s. If $\theta < 0$, we obtain that $ y(\o) \geq  Y(\o,\cdot)$ $ \Pc(\o)$-q.s. Then Theorem \ref{Essup} shows that $ y(\o) \geq  \essup[\Pc]  Y(\o).$ So, $ y(\o) =  \essup[\Pc]  Y(\o),$ a contradiction. The same reasoning applies if $\theta >0.$ Thus $\theta=0$ and 
NA$(\Pc(\o))$ holds true.

\section{Absence of Immediate Profit in a Multiperiod Market Model}
\label{sec:multi}
We consider a multiperiod model with time horizon $T\in \mathbb{N}$ and with  $d$ risky assets and one perfectly liquid num\'eraire which cost is equal to 1. We introduce  a sequence $\left(\Omega_t\right)_{t \in \{1,\ldots,T\}}$  of Polish spaces. Let $1\leq t\leq l\leq T$. We denote by   $\Omega^{t,l}:=\O_{t} \times \dots \times \O_{l}$  and $\Omega^{t}:=\Omega^{1,t}$ (with the convention that $\Omega^{0}$ is reduced to a singleton $\{\o_0\}$). An element of $\Omega^{t,l}$ will be denoted by $\o^{t,l}=(\omega_{t},\dots, \omega_{l})$ for $(\o_{t},\dots,\o_{l}) \in \Omega_{t}\times\dots\times\Omega_{l}$. We will write  $\o^{t}:=\o^{1,t}$. We also define 
$\mathfrak{S}_l(\o^t):=\{\tilde \o^l \in \O^l:\; \tilde \o^t =\o^t\}=\{\o^t\}\times \Omega^{t+1,l}.$ 
To avoid heavy notation we drop the dependency in $\o_{0}$. 

Let $\mathcal{Q}_{1}$ be a nonempty subset of $\mathfrak{P}(\O_{1}).$
 For all $t \in \{1,\ldots,T-1\}$, let $\mathcal{Q}_{t+1}: \Omega^t \twoheadrightarrow \mathfrak{P}(\O_{t+1}),$   where $\mathcal{Q}_{t+1}(\o^{t})$  can be seen as the set of all possible  priors for the $t+1$-th period given the state $\o^{t}$ until time $t$. 
We assume that  $\mathcal{Q}_{t+1}$  satisfies Assumption \ref{anaone}. 
So, the Jankov-von Neumann theorem (see  \cite[Proposition 7.49 p182]{bs}) implies that there exists a $\mathcal{B}_{c}(\O^{t})$-measurable $q_{t+1}: \Omega^{t} \to \mathfrak{P}(\O_{t+1})$ such that $q_{t+1}(\cdot,\o^{t}) \in \mathcal{Q}_{t+1}(\o^{t})$  for all $\o^{t} \in \O^{t}$.  \\
For all $0 \leq t <l \leq T$ let $\mathcal{Q}_{t+1,l}: \Omega^t \twoheadrightarrow \mathfrak{P}\left(\Omega^{t+1,l}\right)$  be defined by
\begin{align*}
 \mathcal{Q}_{t+1,l}(\o^t): & = 
 \{ q_{t+1} \otimes \dots \otimes q_{l}(\cdot,\o^{t}):\;    q_{k+1} \in \mathcal{S}K_{k+1}(\o^t)\\
& 
q_{k+1}(\cdot,\o^{k}) \in \mathcal{Q}_{k+1}(\o^{k}), \;  
 \forall \; \o^{k} \in  \mathfrak{S}_k(\o^t),  \; \forall \; k \in \{t,\ldots,l-1\} \} 
\end{align*}
where for $ k \in \{t,\ldots,l-1\}$  
$\mathcal{S}K_{k+1}(\o^t)$ is a set of stochastic kernels  
such that $q_{k+1}(\cdot,\o^{k})$ is a probability measure on $\mathfrak{P}(\O_{k+1})$ for all $\o^k \in \mathfrak{S}_k(\o^t)$ and 
$\omega^{t+1,k} \mapsto q_{k+1}(A,(\o^{t},\omega^{t+1,k}))$ is $\Bc_c(\Omega^{t+1,k})$-measurable for all $A\in \Bc(\O_{k+1})$. 
Note that if $l=t+1$ this last condition vanishes and $\mathcal{Q}_{t+1,t+1}(\o^t)= \mathcal{Q}_{t+1}(\o^t).$ 
Moreover, using Fubini's theorem (see  \cite[Proposition 7.45 p175]{bs}), we have set for all $\o^t \in \O^t$ 
and $A \in \mathcal{B}_c(\Omega^{t+1,l})$
\begin{align*}
q_{t+1} \otimes \dots \otimes q_{l}(A,\o^{t}): & = 
\begin{multlined}[t][\textwidth-3.7cm]
 \int_{\Omega_{t+1}} \ldots \int_{\Omega_{l}}  1_{A}(\omega_{t+1}, \ldots, \omega_{l}) \\q_{l}(d\omega_{l},(\omega^{t},\omega_{t+1},\ldots,\o_{l-1})) \ldots q_{t+1}(d\omega_{t+1},\omega^{t}).
 \end{multlined}
\end{align*}
For all $t \in \{1,\ldots,T\},$ we also set $\mathcal{Q}^{t}:= \mathcal{Q}_{1,t} \subset \mathfrak{P}\left(\Omega^{t}\right).$ 

Usually in the quasi-sure literature only the  sets $\mathcal{Q}_{t+1}(\o^t)$  and $\mathcal{Q}^{t}$ are  of interest.  For example the global no-arbitrage condition requires
$\mathcal{Q}^{T}{\rm -q.s.}$ inequalities while local no-arbitrage requires $\mathcal{Q}_{t+1}(\o^{t}) {\rm -q.s.}$ ones, see Definition \ref{NAQTARB}. The equivalence 
between both notions holds true for $\o^t$ in a $\mathcal{Q}^{t}$-full measure set, see Theorem \ref{bnlocal}. 
The reason for introducing the set  $\mathcal{Q}_{t+1,T}(\o^t)$ is related to the definition of global AIP in multiperiod market model, see Definition \ref{foire}. Already, in the uni-prior setup (see  \cite[Proposition 3.4]{BCL}), in order 
to get the equivalence between the  global AIP and the one-step $\mbox{AIP}_{t}$ for all $t$, it is necessary 
to define the global AIP  between times $t$ and $T$ for all $t$ and not only between $0$ and $T$ as in the case of global no-arbitrage. 
This implies that conditions holding $\mathcal{Q}_{t+1}(\o^{t})$ for all $t$ between $0$ and $T-1$ are not sufficient, we need $\mathcal{Q}_{t+1,T}(\o^t){\rm -q.s.}$ inequalities.  
The pendent of \cite[Proposition 3.4]{BCL} is established in Theorem \ref{thoNip01}. The fact that we work with $\mathcal{Q}_{t+1,T}(\o^t)$ complexifies a lot the proof of Theorem \ref{thoNip01} and requires advanced tools of measure theory (see Proposition \ref{lelemmebis} and the comments in the appendix). 

As in the one period model, the trading costs for the $d$ risky assets are given by a sequence of real-valued functions $ \Sigma=( \Sigma_t)_{t \in \{0,\ldots,T\}}$ 
where  $\Sigma_t :\O^t  \times 
 \R^{d}  \to   \R$ for all $t \in \{0,\ldots,T\}.$  
Let $ \theta:=\{ \theta_{t}=\left(\theta^{i}_{t}\right)_{i \in \{1,\ldots,d\}}: \,t \in \{0,\ldots,T\}\} \subset \R^d$  and  $ \theta^0:=\{ \theta^0_{t}: \,t \in \{0,\ldots,T\}\} \subset \R.$  The processes $\theta$ and $\theta^0$  represent respectively the
investor's holdings in  
 the $d$ risky assets and in the num\'eraire. 
The interpretation is that buying a portfolio $(\theta^0,\theta) \in  \R^{d+1}$ at time $t$ and in state $\o^t \in \O^t$ costs $ \theta^0+  \Sigma_t (\o^t,  \theta)$.  
We define 
\begin{eqnarray*}
\Sc_{t,T}& := & \{(\theta_t, \ldots, \theta_{T+1}) \in \R^{d(T-t+2)} : \; \theta_t=\theta_{T+1}=0\}\\
\Ac_{t,T}& := & \left\{(\theta_t, \ldots, \theta_{T+1}) \in  \prod_{u=t}^{T+1} L^0\left(\R^d,\Bc_c (\O^{u-1})\right): \; (\theta_t, \ldots, \theta_{T+1}) \in\Sc_{t,T} \right\}.
\end{eqnarray*}

In a multiperiod model, a global IP means that there exists some  time $t\in \{0,\ldots,T-1\}$ such that it is possible  to super-replicate from a negative cost  paid at time $t,$ the claim $0$  at time $T$. Formally,  there is a global IP if there exist some $X \in L^0(\R,\Bc_c (\O^{t}))$ such that $ X \leq 0$  $\Qc^t$-q.s. and some $P \in \Qc^t$ such that $P(X<0)>0$ (i.e. $X \in L_{\Qc^t}^0(\R_-,\Bc_c (\O^{t})) \setminus\{0\}$, see \eqref{eqcaselec2}) and 
some $(\hat \theta_t,  \ldots, \hat \theta_{T+1}) \in \Ac_{t,T}$ such that for all $\o^t$ in a $\Qc^t$-full measure set 
\begin{eqnarray}
\label{dinguerie0}
X(\o^t)-\sum_{u=t}^T \Sigma_u((\o^t,\cdot) , \Delta \hat \theta_{u+1}(\o^t ,\cdot))  \geq 0   \,\, \Qc_{t+1,T}(\o^t){\rm -q.s.} 
\end{eqnarray}
First we rewrite the global IP condition in a more convenient way. For that we introduce for all $t \in \{0,\ldots,T-1\}$, the set $\Pi_{t,T}^{\Sigma} (0):$ 
\begin{eqnarray}
\label{epimulti}
\begin{split}
  \Pi_{t,T}^{\Sigma} (0)  :=  \{ X \in  L^0(\R,\Bc_c (\O^t) ) : \exists 
(\theta_t, \ldots, \theta_{T+1}) \in   \Ac_{t,T}  : \; \\
    X(\o^t) - \sum_{u=t}^T \Sigma_u((\o^t,\cdot) , \Delta \theta_{u+1}(\o^t ,\cdot))  \geq 0 \, \Qc_{t+1,T}(\o^t) {\rm -q.s.} \} \\
      \mbox{ for $\o^t$ in a $\Qc^t$-full measure set} \}. 
\end{split}
\end{eqnarray}
Note that $\Pi_{t,T}^{\Sigma} (0)$ is a set of $\Bc_c (\O^t)$-measurable functions in contrast to $\Pi_{t}^{\Sigma} (0)(\o^t)$ which is a set of numbers. With this definition 
\eqref{dinguerie0} says that there is  a global IP if and only if there exist some $t$ and some $X \in  \Pi_{t,T}^{\Sigma} (0) \cap L_{\Qc^t}^0(\R_-,\Bc_c (\O^{t})) \setminus\{0\}$. Now, 
the  one-step IP between $t$ and $t+1$ is defined so that both notions coincide for a one-period model. 
\begin{definition}
\label{foire}
The global AIP  condition holds true if and only if for all  $t \in \{0,\ldots,T-1\},$ 
$$  \Pi_{t,T}^{\Sigma} (0) \cap  L_{\Qc^t}^0(\R_-,\Bc_c (\O^{t})) \setminus\{0\}= \emptyset.$$
Fix some $t \in \{0,\ldots,T-1\}.$ The one-step $\mbox{AIP}_{t}$  condition holds true  if and only if  
$$  \Pi_{t,t+1}^{\Sigma} (0) \cap  L_{\Qc^t}^0(\R_-,\Bc_c (\O^{t})) \setminus\{0\}=\emptyset.$$
\end{definition}
First we show that under an additional condition on the cost functions at time 0, a global IP have a simpler characterization. Namely, it is equivalent to the existence of some  time $t\in \{0,\ldots,T-1\}$ such that it is possible  to super-replicate on a non-polar set,  from a negative cost  paid at time $t,$ the claim $0$  at time $T$. 
\begin{lemma}
\label{leplus}
Assume that for all $t \in \{0,\ldots,T\},$ $\Sigma_{t}(\cdot,0) \leq 0.$ \\
There is a global IP if and only if there exist $t \in \{0,\ldots,T-1\},$ $\e>0,$ $I^t \in \Bc_c(\O^t)$ and $P \in \Qc^t$ such that $P(I^t )>0,$ 
and some $(\hat \theta_t,  \ldots, \hat \theta_{T+1}) \in \Ac_{t,T}$ such that for all $\o^t \in I^t$
\begin{eqnarray*}
-\e-\sum_{u=t}^T \Sigma_u((\o^t,\cdot) , \Delta \hat \theta_{u+1}(\o^t ,\cdot))  \geq 0   \,\, \Qc_{t+1,T}(\o^t){\rm -q.s.} 
\end{eqnarray*}
\end{lemma}
{\sl Proof.} 
Assume that a global IP exists  and take $X \in L_{\Qc^t}^0(\R_-,\Bc_c (\O^{t})) \setminus\{0\}$ that satisfies \eqref{dinguerie0}. Then, there exists some $P \in \Qc^t$ such that $P(X<0)>0.$ This is equivalent to the existence of some $\e>0$ such $P(X\leq -\e)>0$. 
There exist also some $\O^t_{IP} \in \Bc_c(\O^t)$ of $\Qc^t$-full measure, some $(\hat \theta_t,  \ldots, \hat \theta_{T+1}) \in \Ac_{t,T}$ 
such that for all   $ \o^t \in \O^t_{IP}$
 \begin{eqnarray*}
X(\o^t) - \sum_{u=t}^T \Sigma_u((\o^t,\cdot), \Delta \hat \theta_{u+1}(\o^t ,\cdot))  \geq 0   \,\, \Qc_{t+1,T}(\o^t){\rm -q.s.} 
\end{eqnarray*}
and  we just set $I^t =\{X \leq -\e \} \cap \O^t_{IP}$. \\
For all $ u \in \{t, \ldots,T\},$ $\Sigma_{u}(\cdot,0) \leq 0,$ so $0\in \Pi_{t,T}^{\Sigma} (0).$ The reverse implication is proved choosing $X=-\e 1_{I^t}$ and changing $\hat \theta_{u}$ into $\hat \theta_{u}1_{I^t},$ for all $ u \in \{t+1, \ldots,T\}$ (recall that $\hat \theta_t=\hat \theta_{T+1}=0$).  
$\Box$\\

We want to make the link with the results obtained in Section \ref{onegene}. For that we choose for $t \in \{0,\ldots,T-1\},$ $\O=\O^t$,  $\widetilde \O= \O_{t+1}$, $\Pc=\Qc_{t+1},$  
$\Sigma_0=\Sigma_t$ and $\Sigma_1=\Sigma_{t+1}$. 
So, \eqref{laballe5} becomes 
$$\Phi^{\Sigma}_{t}(\o^t,\o_{t+1},\theta):=\Sigma_t(\o^t,  \theta) +\Sigma_{t+1}(\o^t,\o_{t+1}, - \theta).$$
This leads to the following notations  (see \eqref{laballePhi} and Definition \ref{defpgs}) 
\begin{eqnarray}
\label{pionestep0}
\Pi^{\Sigma}_{t}(0)(\o^t) & : = &  \Pi(\Phi^{\Sigma}_{t})(\o^t) \\
\label{pionestep}
 &= &    \left\{x_t\in \R: \;  \exists\, \theta_{t+1}\in \R^d,\;   x_t-\Sigma_t(\o^t,  \theta_{t+1}) 
 \right.\\  
  \nonumber
  & & \left. 
  -\Sigma_{t+1}(\o^t,\cdot, - \theta_{t+1}) \ge 0 \, \Qc_{t+1}(\o^t){\rm -q.s.}\right\}\\
\label{pionestep1}
\pi^{\Sigma}_{t}(0)(\o^t) & : = &  \inf \Pi^{\Sigma}_{t}(0)(\o^t). 
\end{eqnarray}

We want to find the relation between the one-step $\mbox{AIP}_{t}$  condition and  \\
$\mbox{AIP}(\Qc_{t+1}(\cdot))$ (see Definition \ref{defipone}).  For that, we introduce the set $F^t \subset \O^t$ where $\mbox{AIP}(\Qc_{t+1}(\o^t))$ holds true. 
\begin{eqnarray}
\label{fifi}
\scalemath{0.98}{
F^t :  =  \left\{\o^t \in \O^t:\; \mbox{AIP}(\Qc_{t+1}(\o^t)) \mbox{ holds true}\right\} =
\{\o^t \in \O^t:\; \pi^{\Sigma}_t(0)(\o^t) \geq 0\}.}
\end{eqnarray}
\begin{proposition}
\label{ailocdet}
Fix some $t \in \{0,\ldots,T-1\}.$ 
Assume that Assumption \ref{anaone}  holds true for $\Qc_{t+1}$ and that $(\o^t,\o_{t+1}) \mapsto \Phi^{\Sigma}_{t}(\o^t,\o_{t+1},\theta)$ is usa for all $\theta \in \R^d$.\\
If Assumption \ref{hypophi2} holds true for $\Phi^{\Sigma}_{t},$ then $F^t \in \Ac(\O^t)$. \\
Assume now that Assumption \ref{hypophi3} holds true for $\Phi^{\Sigma}_{t}$ and that for all $\o^t \in \Omega^t,$ $\theta \mapsto \Phi^{\Sigma}_{t}(\o^t,\cdot,\theta)$ is $\Qc_{t+1}(\o^t)$-q.s. lsc  and that $\Phi^{\Sigma}_{t}(\o^t,\cdot,0) \le 0$ $\Qc_{t+1}(\o^t)$-q.s.   
The following conditions are equivalent :
\begin{enumerate}
\item The set $ F^t:=\left\{\o^t \in \O^t:\; \mbox{AIP}(\Qc_{t+1}(\o^t)) \mbox{ holds true}\right\} $ is of $\Qc^t$-full measure. 
\item The one-step $\mbox{AIP}_{t}$  condition holds true. 
\end{enumerate}
\end{proposition}


\noindent {\sl Proof.}
As  $\pi^\Sigma_{t}(0)(\cdot)$ is usa (see Proposition \ref{zorro2}), \eqref{fifi} shows that $F^t\in \Ac(\O^t).$ 
Now  Definition \ref{defipone} implies that 
\begin{eqnarray}
\label{greveg}
F^t= \left\{\o^t \in \O^t:\; \Pi^{\Sigma}_{t}(0)(\o^t)\cap (-\infty,0)=\emptyset \right\}.
 \end{eqnarray} 
Assume that 1. holds true. 
Let $X \in  \Pi_{t,t+1}^{\Sigma} (0)\cap  L^0_{\Qc^t}(\R_-,\Bc_c (\O^{t})) \setminus \{0\}.$ So, $X \in  L^0(\R,\Bc_c (\O^t) ),$ there exists some $P\in \Qc^t$ such that $P(X<0)>0,$ there exist some $\Qc^{t}$-full measure set $G^t$ and some $\theta \in  L^0(\R^d,\Bc_c (\O^t) )$ such that  for all  $\o^t \in G^t,$   $X(\o^t) \leq 0$ and 
$$
   X(\o^t) - \Sigma_t(\o^t,  \theta(\o^t))  -\Sigma_{t+1}((\o^t,\cdot) , -\theta(\o^t))\geq 0 \, \Qc_{t+1}(\o^t) {\rm -q.s.}$$
So,  $X(\o^t) \in  \Pi^{\Sigma}_{t}(0)(\o^t)\cap (-\infty,0]$ for all  $\o^t \in G^t$. 
Take now $\o^t \in F^t\cap G^t$. 
If $X(\o^t)<0$, \eqref{greveg} gives a contradiction. Thus, $X(\o^t)=0$ for all $\o^t \in F^t\cap G^t,$ which is of $\Qc^{t}$-full measure, since 1. holds true. As $P(X<0)>0,$  we get a contradiction and 2. is proved. 

Assume now that 2. holds true. Recalling \eqref{fifi}, 
we have to prove that for all $P\in  \Qc^t$, $P(\pi^{\Sigma}_t(0)(\cdot)\geq 0)=1$. Fix some $P\in  \Qc^t.$ Using \eqref{vache}, we get that 
$$\{\o^t \in \O^t:\; \pi^{\Sigma}_{t}(0)(\o^t)<0\}=\cup_{(\theta,\epsilon) \in \QQ^d \times \QQ_+}  \{\o^t \in \O^t:\; \f^\Sigma_t(\o^t,\theta)\leq -\e\},$$
where $\f^\Sigma_t:\O^t \times \R^d \to \R \cup \{+\infty\}$ is the quasi-sure essential supremum of $\Phi^\Sigma_t ( \cdot,\cdot,\theta)$ and has been defined in Proposition \ref{zorro1}
$$\f^\Sigma_t(\o^t,\theta):= \essup[\Qc_{t+1}] \Phi^\Sigma_t ( \cdot,\cdot,\theta) (\o^t).$$
Then, 
\begin{eqnarray}
\label{lafin}
P(\{\pi^{\Sigma}_{t}(0)(\cdot)<0\}) & \leq  & \sum_{\e \in \Q_+, \theta \in \Q^d}  P\left(\f^\Sigma_t(\cdot,\theta)\leq -\e\right).
 \end{eqnarray}
Recalling \eqref{eqflash} and \eqref{pionestep0}, for all $\o^t \in \O^t$ and $\theta \in \R^d,$   
$\f^\Sigma_t(\o^t ,\theta) \in \Pi(\Phi^{\Sigma}_{t})(\o^t )=\Pi^{\Sigma}_t(0)(\o^t ).$ 
Moreover,  for all $\theta \in \R^d$, Theorem \ref{Essup} shows that $\f^\Sigma_t(\cdot ,\theta)$ is usa and thus $\Bc_c (\O^t)$-measurable. 
So,  $$\f^\Sigma_t(\cdot ,\theta) \in L^0_{\Qc^t}(\Pi^{\Sigma}_t(0),\Bc_c (\O^t)).$$
Using 
\eqref{laballebis} in Proposition \ref{zorro2}, there exists  $\hat \theta \in L^0(\R^d,\Bc_c (\O^t )),$ such that 
for $\o^t$ in a $\Qc^t$-full measure set 
$$    \f^\Sigma_t(\o^t ,\theta) \ge \Phi^{\Sigma}_{t} (\o^t,\cdot,\hat \theta (\o^t)) \, \Qc_{t+1}(\o^t) {\rm -q.s.}$$
and $\f^\Sigma_t(\cdot,\theta) \in \Pi_{t,t+1}^{\Sigma} (0)$ for all $\theta \in \R^d.$   
Assume now that there exist some $\e >0$ and $\theta \in \R^d$ such that $ P\left(\f^\Sigma_t(\cdot,\theta)\leq -\e\right)>0$. 
As $0\in  \Pi_{t,t+1}^{\Sigma} (0)$ by assumption, we get that 
$$\f^\Sigma_t(\cdot,\theta) \one_{\{\f^\Sigma_t (\cdot,\theta)\leq -\e\}} (\cdot)\in  \Pi_{t,t+1}^{\Sigma} (0)\cap  L^0_{\Qc^t}(\R_-,\Bc_c (\O^t)) \setminus \{0\},$$ 
a contradiction to 2. 
So, \eqref{lafin} shows that 
$P(\{\pi^{\Sigma}_{t}(0)(\cdot)\geq 0\})=1$ which concludes the proof of 1. 
$\Box$\\

Now, we show the equivalence between 
the locals and the global AIP conditions. For that we need the following assumption on the cost functions. 

\begin{assumption}
\label{costfun}
The cost functions  
 $ \Sigma_t :\O^t  \times 
 \R^{d}  \to \R$ satisfy for all $t \in \{0,\ldots,T\}$  the following conditions: 
 \begin{enumerate}
 \item  $\theta \mapsto  \Sigma_t(\o^t,\theta)$ is lsc for $\o^t$ in a $\Qc^t$-full measure set. 
 \item $\theta \mapsto  \Sigma_t(\o^t,\theta)$ is sublinear for $\o^t$ in a $\Qc^t$-full measure set, i.e. for all $\theta,\,\theta\rq{} \in \R^d$
 $$\Sigma_t(\o^t,\theta+\theta\rq{}) \geq \Sigma_t(\o^t,\theta) +\Sigma_t(\o^t,\theta\rq{}).$$
 \item   $(\o^t, \theta) \mapsto  \Sigma_t(\o^t, \theta)$ is $\Bc(\O^t)\otimes \Bc(\R^{d})$-measurable.
 \item $ \Sigma_t(\cdot, 0) \leq 0.$
 \item Assumption \ref{hypophi3} holds true for $\Phi^{\Sigma}_{t}.$
 \end{enumerate}
\end{assumption}
The sublinearity of $\Sigma_t$ says that it costs more to buy the portfolios $\theta$ and $\theta\rq{}$ together than separately. This is coherent with the way  the trading algorithms work: it is cheaper to execute orders in small blocks than to sell the whole portfolio in one unique order. \\
Note that Condition  3. implies that $\o^t \mapsto  \Sigma_t(\o^t, \theta)$ is  $\Bc(\O^t)$-measurable and thus  usa for all $\theta \in \R^d,$ see \cite[Lemma 4.48 p152]{Hitch}. 
\begin{remark}
\label{lumiere}
Assume that  $ \Sigma_t(\o^t, \cdot): \, \theta \mapsto \Sigma_t(\o^t, \theta)$ is  additive, i.e. that for all $\theta,\, \theta\rq{} \in \R^d$, 
$$\Sigma_t(\o^t, \theta +\theta\rq{})=\Sigma_t(\o^t, \theta)+\Sigma_t(\o^t, \theta\rq{}).$$
As $ \Sigma_t(\o^t, \cdot)$ is $\Bc(\R^d)$-measurable, it  is continuous (see for example \cite{BA20}) and thus linear. So, there exists a $\mathbb{R}^{d}$-valued process $ S:=\left\{S_{t}:\, t \in \{0,\ldots,T\}\right\}$ such that 
\begin{eqnarray}
\label{pinot}
 \Sigma_t(\o^t,  \theta)=   \theta  S_t(\o^t).
\end{eqnarray} 
Moreover, $S_t$ will inherit the measurability properties of $\Sigma_t( \cdot,\theta)$ and is thus $\Bc(\O^t)$-measurable. The assumption that $S_t$ is Borel measurable is postulated in the frictionless literature on quasi-sure finance, see \cite{BN}. 
\end{remark}

\begin{theorem}\label{thoNip01} 
Assume that Assumption \ref{anaone}  holds true for $\Qc_{t+1}$ for all  $t \in \{0,\ldots,T-1\}.$ Assume also that Assumption \ref{costfun} holds true. \\
The global AIP condition holds true if and only if the one-step $\mbox{AIP}_{t}$  condition holds true for all $t\in \{0,\ldots,T-1\}.$ 
\end{theorem}
{\sl Proof. }
The reverse implication is proved in Lemma \ref{lemcool}  below. We prove the direct implication by contraposition and 
assume that for some $t\in \{0,\ldots,T-1\},$ there is a one-step  $\mbox{IP}_{t}.$ 
Recalling \eqref{dinguerie0}, there exists some $\hat \theta_{t+1} \in L^0(\R^d,\Bc_c (\O^t))$,   $P\in \Qc^t$, $X \in L^0(\R,\Bc_c (\O^{t}))$ such that 
$X\leq 0 \,\Qc^t{\rm -q.s.},$ 
$P(X<0)>0$ and   for all $\o^t$ in a $\Qc^t$-full measure set 
\begin{eqnarray*}
X(\o^t)- \Sigma_t(\o^t, \hat \theta_{t+1}(\o^t ))  + \Sigma_{t+1}((\o^t,\cdot) , - \hat  \theta_{t+1}(\o^t ))\geq 0   \,\, \Qc_{t+1}(\o^t){\rm -q.s.} 
\end{eqnarray*}
This last inequality also holds true $\Qc_{t+1,T}(\o^t){\rm -q.s.}$ by Fubini theorem.  
We set $\hat \theta_{t}=\hat \theta_{t+2}=\cdots=\hat \theta_{T+1}=0.$ Then, $(\hat \theta_t,\ldots, \hat \theta_{T+1})\in \Ac_{t,T}.$ As 
 $ \Sigma_u(\cdot, 0) \leq 0 $ for $u \in \{t+2,\ldots,T\},$ we get that 
for all $\o^t$ in a $\Qc^t$-full measure set 
\begin{eqnarray*}
X(\o^t)-\sum_{u=t}^T \Sigma_u((\o^t,\cdot) , \Delta \hat \theta_{u+1}(\o^t ,\cdot))  \geq 0   \,\, \Qc_{t+1,T}(\o^t){\rm -q.s.} 
\end{eqnarray*}
and the absence of global IP fails. 
$\Box$

 \begin{lemma}\label{lemcool} 
 Assume that Assumption \ref{anaone}  holds true for $\Qc_{t+1}$ for all  $t \in \{0,\ldots,T-1\}.$ Assume also that Assumption \ref{costfun} holds true. \\ 
Fix $t \in \{0,\ldots,T-1\}.$ 
Assume that the one-step $\mbox{AIP}_{s}$  condition holds true for all $s\in \{t,\ldots,T-1\}.$ Then,  
\begin{eqnarray}
\label{graal}
 \Pi_{t,T}^{\Sigma} (0)\cap  L_{\Qc^t}^0(\R_-,\Bc_c (\O^{t}))\setminus\{0\}=\emptyset.
 \end{eqnarray}
\end{lemma}
{\sl Proof.}
If $t=T-1$, the result is immediate as  $\mbox{AIP}_{T-1}$ is defined in 
Definition \ref{foire} by
$$ \Pi_{T-1,T}^{\Sigma} (0)\cap  L_{\Qc^{T-1}}^0(\R_-,\Bc_c (\O^{T-1}))\setminus\{0\}=\emptyset.$$ 
Now, assume that $t=T-2$ and let us prove \eqref{graal} for $t=T-2$.  For that we choose $X \in  L^0(\R,\Bc_c (\O^{T-2}))$ such that $X(\cdot) \leq 0$ $\Qc^{T-2}$-q.s. and such that there exists some $P\in \Qc^{T-2}$ verifying  
$P(X<0)>0$. We also assume that  $X \in    \Pi_{T-2,T}^{\Sigma} (0).$ We will show that 
 $X \geq 0$ $\Qc^{T-2}$-q.s. which provides the desired contradiction and shows \eqref{graal} for $t=T-2.$\\
Recalling \eqref{epimulti}, there exist $\theta_{T-1} \in L^0(\R^d,\Bc_c (\O^{T-2})),$ $\theta_{T} \in L^0(\R^d,\Bc_c (\O^{T-1}))$ and $\O^{T-2}_{IP} \in \Bc(\O^{T-2})$ a  $\Qc^{T-2}$-full measure set, such that for all $\o^{T-2} \in \O^{T-2}_{IP},$ 
\begin{eqnarray}
\label{feur}
\begin{split}
X(\o^{T-2})-\Sigma_{T-2}(\o^{T-2},\theta_{T-1}(\o^{T-2})) -
\Sigma_{T-1}((\o^{T-2},\cdot), \Delta  \theta_{T}(\o^{T-2},\cdot))  \\
- \Sigma_{T}((\o^{T-2},\cdot),- \theta_{T}(\o^{T-2},\cdot))
 \geq 0   \,\, \Qc_{T-1,T}(\o^{T-2}){\rm -q.s.} 
 \end{split}
\end{eqnarray}  
We want to transform \eqref{feur} above which holds true $\Qc_{T-1,T}(\o^{T-2})$-q.s. for  $\o^{T-2} \in \O^{T-2}_{IP}$   into an inequality which holds true $\Qc_{T}(\o^{T-1}){\rm -q.s.} $ for  $\o^{T-1} \in \tilde \Omega^{T-1},$  where $\tilde \Omega^{T-1} \in \Bc_c(\O^{T-1})$ and is a $\Qc^{T-1}$-full measure set. For that we want to apply 
 Proposition \ref{lelemmebis} to 
\begin{eqnarray*}
\Phi (\o^{T-2},\o_{T-1},\o_{T}) =X(\o^{T-2})-\Sigma_{T-2}(\o^{T-2},\theta_{T-1}(\o^{T-2}))  \\
-
\Sigma_{T-1}((\o^{T-2},\o_{T-1}), \Delta  \theta_{T}(\o^{T-2},\o_{T-1})) \\ - \Sigma_{T}((\o^{T-2},\o_{T-1},\o_{T}),- \theta_{T}(\o^{T-2},\o_{T-1})).
\end{eqnarray*} 
The function  $\Phi$ is $\mathcal{B}_{c}(\O^{T-1} )\otimes \mathcal{B}(\O_{T})$-measurable.
Indeed, $X \in  L^0(\R,\Bc_c (\O^{T-2})),$ $\theta_{i+1} \in L^0(\R^d,\Bc_c (\O^{i}))$ for $ i \in \{T-2, T-1\}$ and  $(\o^i, \theta) \mapsto  \Sigma_i(\o^i,\theta)$ is $\Bc(\O^i)\otimes \Bc(\R^{d+1})$-measurable for $T-2\leq i \leq T.$  Moreover, \eqref{feur} asserts that $\Phi (\o^{T-2},\cdot) \geq 0$ $\Qc_{T-1,T}(\o^{T-2}){\rm -q.s.}$ for all 
$\o^{T-2} \in \O^{T-2}_{IP}.$ So, denoting by $\tilde \Omega^{T-1}$ the graph of the set-valued mapping $A^{T-1,T-1}$ of Proposition \ref{lelemmebis}, 
 $\tilde \Omega^{T-1} \in \Bc_c(\O^{T-1})$ and  is of $\Qc^{T-1}$-full measure and  for all
$\o^{T-1}\in \tilde \Omega^{T-1}$  
\begin{eqnarray*} 
\nonumber X(\o^{T-2})-\Sigma_{T-2}(\o^{T-2},\theta_{T-1}(\o^{T-2})) -
\Sigma_{T-1}(\o^{T-1}, \Delta  \theta_{T}(\o^{T-1})) \\ 
- \Sigma_{T}((\o^{T-1},\cdot),- \theta_{T}(\o^{T-1}))\geq 0\;\Qc_{T}(\o^{T-1}){\rm -q.s.}
\end{eqnarray*}
As $\Sigma_{T-1}$ is {$\Qc^{T-1}$} {sub}linear, there exists $\bar \Omega^{T-1} \in \Bc_c(\O^{T-1})$ which is of $\Qc^{T-1}$-full measure and such that for all
$\o^{T-1}\in \bar \Omega^{T-1}$  
\begin{eqnarray*} 
\scalemath{0.97}{\Sigma_{T-1}(\o^{T-1}, \Delta  \theta_{T}(\o^{T-1})) \ge \Sigma_{T-1}(\o^{T-1},   \theta_{T}(\o^{T-1}))+ \Sigma_{T-1}(\o^{T-1}, - \theta_{T-1}(\o^{T-2})).}
\end{eqnarray*} 
So, for all    $\o^{T-1} \in \widetilde{\O}^{T-1} \cap \bar \Omega^{T-1},$ which is a  $\Qc^{T-1}$-full measure set, we obtain that $\Qc_{T}(\o^{T-1}){\rm -q.s.}$
\begin{align*} X(\o^{T-2})  -\Sigma_{T-2}(\o^{T-2},\theta_{T-1}(\o^{T-2})) -\Sigma_{T-1}(\o^{T-1}, - \theta_{T-1}(\o^{T-2})) &\\
-\Sigma_{T-1}(\o^{T-1}, \theta_{T}(\o^{T-1})) -
\Sigma_{T}((\o^{T-1},\cdot),- \theta_{T}(\o^{T-1}))  \geq 0.&
\end{align*}
Let 
\begin{eqnarray*}
\hat \Omega^{T-2} & :=  &\O^{T-2}_{IP} \cap \{X \leq 0\} \cap F^{T-2} \\
\widehat{\O}^{T-1}& :=  &\widetilde{\O}^{T-1} \cap \bar \Omega^{T-1} \cap   (\hat \O^{T-2}\times \O_{T-1})\cap F^{T-1},
\end{eqnarray*}
see \eqref{fifi} for the definition of $F^{T-2}$ and $F^{T-1}.$ Then for $s \in \{T-1,T-2\}$ $\hat \Omega^{s}$ belongs to $\Bc_c(\O^{s})$ and is of  $\Qc^{s}$-full measure as $\mbox{AIP}_{s}$ holds true, see Proposition \ref{ailocdet}. 
For all
$\o^{T-1}\in \widehat{\O}^{T-1},$ recalling \eqref{pionestep} and \eqref{pionestep1}, we get that 
\begin{eqnarray*}
\scalemath{0.9}{ X(\o^{T-2})-\Sigma_{T-2}(\o^{T-2},\theta_{T-1}(\o^{T-2})) -\Sigma_{T-1}(\o^{T-1}, - \theta_{T-1}(\o^{T-2}))
\geq \p_{T-1}^\Sigma(0)(\o^{T-1}).}
\end{eqnarray*}
As $\o^{T-1}\in F^{T-1},$ $\p_{T-1}^\Sigma(0)(\o^{T-1}) \geq 0$  using \eqref{fifi}. 
So, 
$$X(\cdot) -\Sigma_{T-2}(\cdot,\theta_{T-1}(\cdot) ) - \Sigma_{T-1}(\cdot,-\theta_{T-1}(\cdot) )\geq 0 \; \Qc^{T-1}{\rm -q.s.}$$ 
Now, we use Lemma \ref{lelemme} with $\O=\O^{T-2}$, $\tilde \O=\O_{T-1},$ $\Rc=\Qc^{T-2},$ $\Pc=\Qc_{T-1}$,  
and \begin{eqnarray*}\overline B=\{(\o^{T-2},\o_{T-1}):\, X(\o^{T-2}) -\Sigma_{T-2}(\o^{T-2},\theta_{T-1}(\o^{T-2}) ) - \\
\Sigma_{T-1}((\o^{T-2},\o_{T-1}),-\theta_{T-1}(\o^{T-2}) )\geq 0 \}.\end{eqnarray*}
Indeed, as before 
$\overline B \in \Bc_c(\O^{T-2}) \otimes  \Bc(\O_{T-1})$ and  we obtain that 
\begin{eqnarray*}
B=\{\o^{T-2} \in \O^{T-2}: \;X(\o^{T-2})-\Sigma_{T-2}(\o^{T-2},\theta_{T-1}(\o^{T-2}))-\\
\Sigma_{T-1}((\o^{T-2},\cdot), -\theta_{T-1}(\o^{T-2}))\geq  0 \,\Qc_{T-1}(\o^{T-2}){\rm -q.s.}\}\end{eqnarray*}
belongs to $\Bc_c(\O^{T-2})$ and is of $\Qc^{T-2}$-full measure. \\
Using \eqref{pionestep} and \eqref{pionestep1} as before, we obtain for all $\o^{T-2} \in  B$ that 
 $X(\o^{T-2})\geq \pi^{\Sigma}_{T-2}(0)(\o^{T-2}).$
 Thus, choosing  $\o^{T-2} \in \hat \O^{T-2} \cap B,$ we get that  $$X(\o^{T-2})\geq \pi^{\Sigma}_{T-2}(0)(\o^{T-2})\geq 0$$  as $\hat \O^{T-2}\subset F^{T-2}$ and 
 \eqref{fifi} holds true. 
  We have finished the proof of  
 $X(\cdot) \geq 0$  $\Qc^{T-2}{\rm -q.s.}$ We can continue the proof in a similar way. $\Box$\\

Finally, in a frictionless market, we make the comparison with the quasi-sure no-arbitrage condition of \cite{BN}. Assume that \eqref{pinot} holds true, i.e. that for all $t \in \{0,\ldots,T\}$ 
$
 \Sigma_t(\o^t,  \theta)=   \theta  S_t(\o^t), 
$
where  $S:=\left\{S_{t}: \, t \in \{0,\ldots,T\}\right\}$ is a 
$\mathbb{R}^d$-valued process. For 
 all $t \in \{0,\ldots,T\}$, $S_{t}=\left(S^i_t\right)_{i \in \{1,\ldots,d\}}$ represents the  market price of $d$ risky assets at time $t$. Note that as mention in Remark \ref{lumiere}, it \rq{}s among to say that $ \Sigma_t$ is additive. 
We assume that 
$S_t \in  L^0(\R^d,\Bc (\O^{t})).$  Using again the notation $\Delta
S_t:=S_t-S_{t-1},$ we obtain that 
$\mathcal{Q}^{T} {\rm -q.s.}$ for all $\theta=(\theta_0, \ldots, \theta_{T+1}) \in \Ac_{0,T} $ 
\begin{eqnarray*}
x-\sum_{t=0}^{T}\Sigma_{t}(\cdot, \Delta \theta_{t+1}( \cdot))    & = & x - \sum_{t=0}^{T}\Delta \theta_{t+1}( \cdot)S_t( \cdot)\\
&  = & 
x+\theta_{T}( \cdot)S_T( \cdot)-\theta_{T}( \cdot)S_{T-1}( \cdot) +\theta_{T-1}( \cdot)S_{T-1}( \cdot) \\
 & & +\ldots -\theta_{2}( \cdot)S_{1}( \cdot) +\theta_{1}( \cdot)S_{1}( \cdot) - \theta_{1}( \cdot)S_{0}( \cdot)\\
& = & x+ \sum_{t=1}^{T} \theta_{t}( \cdot)\Delta S_t( \cdot)=:V_{T}^{0,\theta}.  
\end{eqnarray*} 
\begin{definition}
\label{NAQTARB}
The $NA(\mathcal{Q}^{T})$ condition holds true if
$V_{T}^{0,\theta} \geq 0 \; \mathcal{Q}^{T}{\rm -q.s.}$  for some $\theta  \in \Ac_{0,T}$ implies that $ V_{T}^{0,\theta}   = 0 \;\mathcal{Q}^{T}\mbox{-q.s. }$\\
For $\o^{t} \in \O^{t}$ fixed, the local version of $NA(\mathcal{Q}^{T}),$ denoted by  $NA(\mathcal{Q}_{t+1}(\o^{t})),$  holds true if 
$
h\Delta S_{t+1}(\o^{t},\cdot) \geq 0 \; \mathcal{Q}_{t+1}(\o^{t}) {\rm -q.s.}$  for some $h \in \mathbb{R}^{d}$ implies that  $h\Delta S_{t+1}(\o^{t},\cdot) = 0 \;\mathcal{Q}_{t+1}(\o^t){\rm -q.s.}$ 
\end{definition}
We recall the following equivalence between no-arbitrage, local no-arbitrage and relative interior of the quasi-sure support. 
\begin{theorem}
\label{bnlocal}
 Assume that Assumption \ref{anaone}  holds true for $\Qc_{t+1}$ and that $\mathcal{Q}_{t+1}$ is convex-valued for all  $t \in \{0,\ldots,T-1\}.$  
Assume that  for all $t \in \{0,\ldots,T\},$ 
 $\Sigma_t(\o^t,  \theta)=   \theta  S_t(\o^t)
$
where 
$S_t \in  L^0(\R^d,\Bc (\O^{t})).$ 
The following  statements are equivalent.\\
\noindent 1. The $NA(\mathcal{Q}^{T})$ condition holds true.\\
\noindent 2. For all $t \in \{0,\ldots,T-1\},$  the $NA(\mathcal{Q}_{t+1}(\cdot))$ condition holds true $\mathcal{Q}^{t}{\rm -q.s.}$\\
\noindent 3.  For all $t \in \{0,\ldots,T-1\},$  $0 \in {\rm ri \, conv}\,{\rm supp}_{\mathcal{Q}_{t+1}}\Delta S_{t+1}(\cdot)$  $\mathcal{Q}^{t}{\rm -q.s.}$
\end{theorem}
{\sl Proof.}
The proof of the equivalence between 1. and 2. is provided 
in \cite[Theorem 4.5]{BN}, while the equivalence between 2. and 3.  is proved in \cite[Theorem 3.23]{BC19}. 
$\Box$\\

We now provide a similar theorem for AIP. 
\begin{theorem}
\label{aiplocal}
 Assume that Assumption \ref{anaone}  holds true for $\Qc_{t+1}$ for all  $t \in \{0,\ldots,T-1\}.$  
Assume that  for all $t \in \{0,\ldots,T\},$ 
 $\Sigma_t(\o^t,  \theta)=   \theta  S_t(\o^t)
$
where 
$S_t \in  L^0(\R_+^d,\Bc (\O^{t})).$ 
The following  statements are equivalent.\\
\noindent 1. The global AIP  condition holds true.\\
\noindent 2. For all $t \in \{0,\ldots,T-1\},$  the one step $AIP_t$ holds true.\\
\noindent 3. For all $t \in \{0,\ldots,T-1\},$  the $\mbox{AIP}(\Qc_{t+1}(\cdot) )$ condition holds true $\mathcal{Q}^{t}{\rm -q.s.}$ \\
\noindent 4.  For all $t \in \{0,\ldots,T-1\},$  $0 \in {\rm conv}\,{\rm supp}_{\mathcal{Q}_{t+1}}\Delta S_{t+1}(\cdot)$  $\mathcal{Q}^{t}{\rm -q.s.}$
\end{theorem}
{\sl Proof.}
As $\Sigma_t(\o^t,  \theta)=   \theta  S_t(\o^t) 
$
and 
$S_t \in  L^0(\R_+^d,\Bc (\O^{t})),$ 1., 2., 3. and 4. in Assumption \ref{costfun} are clear. Then, 5. follows from Lemma \ref{unifcont}. 
So, Theorem  \ref{thoNip01} applies and gives the equivalence between 1. and 2.  Proposition \ref{ailocdet} shows the equivalence between 2. and 3. and Corollary 
\ref{NGDtwo} the one between 3. and 4. 
$\Box$\\
Thus, it is clear that $NA(\mathcal{Q}^{T})$ implies the global AIP condition. Moreover the equivalence holds true if \eqref{jacshiCNS} below is satisfied. 
\begin{proposition}
\label{NAAIP}
 Assume that Assumption \ref{anaone}  holds true for $\Qc_{t+1}$ and that $\mathcal{Q}_{t+1}$ is convex-valued for all  $t \in \{0,\ldots,T-1\}.$  
Assume that  for all $t \in \{0,\ldots,T\},$ 
 $\Sigma_t(\o^t,  \theta)=   \theta  S_t(\o^t) 
$
where
$S_t \in  L^0(\R_+^d,\Bc (\O^{t})).$ 
The global $AIP$  and the $NA(\mathcal{Q}^{T})$ conditions are equivalent if and only if for all $t \in \{0,\ldots,T-1\},$
\begin{eqnarray}
\label{jacshiCNS}
\scalemath{0.97}{
0 \in {\rm ri \, conv} \,{\rm supp}_{\mathcal{Q}_{t+1}}\Delta S_{t+1}(\cdot) \; \mathcal{Q}^{t} {\rm -q.s.} \Leftrightarrow
0 \in {\rm  conv}\,{\rm supp}_{\mathcal{Q}_{t+1}}\Delta S_{t+1}(\cdot) \; \mathcal{Q}^{t}{\rm -q.s.}
}
\end{eqnarray}
\end{proposition}

\section{Appendix}
 \appendix\normalsize
%

Recall that  $\O$ and $\widetilde{\O}$ are two Polish spaces and $\Pc : \Omega \twoheadrightarrow \mathfrak{P}(\widetilde{\O})$ is a nonempty set-valued mapping. Recall also that  
$\mathcal{S}K$ is a set of stochastic kernels  
such that $q(\cdot,\o)$ is a probability measure in $\mathfrak{P}(\widetilde{\O})$ for all $\o \in \O$ and 
$\o \mapsto q(A,\o)$ is $\Bc_c(\Omega)$-measurable for all $A\in \Bc(\widetilde{\O})$. 
Let $\Rc \subset \mathfrak{P}({\O}),$ we set 
\begin{eqnarray*}
 \mathcal{Q}:=\{ R \otimes q:\;  R \in \Rc,\;  q \in \mathcal{S}K,
 q(\cdot,\o) \in \Pc(\o)\;  \;  \forall \; \o \in \Omega\}.
\end{eqnarray*} 
In the quasi-sure literature, it is necessary to prove that if some function $\Phi:\Omega \times \tilde{\Omega} \to \R$ satisfies that $\Phi \geq 0$ $ \mathcal{Q}$-q.s., then there exists some $\Rc$-full measure set such that for all $\o$ in this  set, $\Phi(\o,\cdot) \geq 0$ $ \mathcal{Q}(\o)$-q.s. 
When $\Phi$ is the difference of a $\mathcal{B}_{c}(\O)\otimes \mathcal{B}(\widetilde \O)$-measurable function with a usa one, 
this is proved in \cite[Proof of Lemma 4.10, pp.846-848]{BN}, see also \cite[(B.4)]{COW}.   
We prove in Lemma \ref{lelemme} a similar result but for general set $\overline{B} \in \mathcal{B}_{c}(\O)\otimes \mathcal{B}(\widetilde \O)$ and not only for 
 $\overline{B}=\{\Phi  \geq 0\}.$ 
 For that  we introduce for some Polish space $E$ and  some paving $\mathcal{J}$ (i.e.  a nonempty collection of subsets of $E$ containing the empty set),  $\mathfrak{A}(\mathcal{J})$  the set of all nuclei of Suslin Scheme on $\mathcal{J}$ (see \cite[Definition 7.15 p157]{bs}). 
 What is really new is that in Proposition \ref{lelemmebis}  we are able to go beyond the one-step case, i.e.  from $\Qc^T{\rm -q.s.}$ inequality to  $\Qc_{T}(\o^{T-1}){\rm -q.s.}$ one for all $\o^{T-1}$ in a   $\Qc^{T-1}$-full measure set.  
We show that if for all $\o^t$ in a  $\Qc^t$-full measure set  
 $\Phi(\o^t,\cdot) \geq 0\;\Qc_{t+1,T}(\o^t){\rm -q.s.}$ then $\Phi(\o^{T-1},\cdot) \geq 0\;\Qc_{T}(\o^{T-1}){\rm -q.s.}$ for all $\o^{T-1}$ in a   $\Qc^{T-1}$-full measure set. This is proved  assuming only that ${\rm gph} \,   \Qc_{T}$ is an analytic set  on $\O^{T-1} \times \mathfrak{P} (\O_{T})$  and not that all  ${\rm gph}\, \Qc_{l+1,T}$  are  analytic sets  on $\O^{l} \times \mathfrak{P} (\O^{l+1,T})$ with $l\in \{t+1,T-1\}$
\begin{lemma}
\label{lelemme}
 Assume that ${\rm gph} \,   \Pc \in \mathfrak{A} \left(\mathcal{B}_{c}(\O)\otimes \mathcal{B}(\mathcal{P}(\widetilde \O))\right)$.   
Let $\overline{B} \in \mathcal{B}_{c}(\O)\otimes \mathcal{B}(\widetilde \O)$. 
  For $\o \in \O$, we denote by $\overline{B}_{\o}$ the section of $\overline{B}$ along $\o$, i.e. $\overline{B}_{\o}:=\{\widetilde \o \in \widetilde \O: \; (\o,\widetilde \o) \in \overline{B}\}$. Then,
\begin{eqnarray}
\label{BB}B:=\left\{\o \in \Omega: \;  q\left(\overline{B}_{\o}\right)=1, \; \forall \; q \in \Pc(\o)\right\} \in  \mathcal{B}_{c}(\O).
\end{eqnarray}
 If $\overline{B}$ is a $\Qc$-full measure set, then $B$ is a $\Rc$-full measure set.
\end{lemma}
{\sl Proof.}
First, as $\overline{B} \in  \mathcal{B}_{c}(\O)\otimes \mathcal{B}(\widetilde \O)$, \cite[Lemma 4.46 p151]{Hitch} shows that $\overline{B}_{\o}\in \mathcal{B}(\widetilde \O)$ for all $\o \in \O$.
Remark that $B=\{\Lambda \geq 1\}$, where 
$$\Lambda(\o):=\inf_{q \in \Pc(\o)} q\left(\overline{B}_{\o}\right).$$
 We claim that $\Lambda$ is $\mathcal{B}_{c}(\O)$-measurable. 
Indeed, $(\o,\widetilde \o)  \mapsto 1_{\overline{B}_{\o}}(\widetilde \o)=1_{\overline{B}}(\o,\widetilde \o) $ is  $\mathcal{B}_{c}(\O)\otimes \mathcal{B}(\widetilde \O)$-measurable. So, a measurable class  argument shows that 
$$\Omega \times \mathcal{P}(\widetilde \O)  \ni (\o,q) \mapsto\lambda(\o,q):= q\left(\overline{B}_{\o}\right)$$ is $\mathcal{B}_{c}(\O)\otimes \mathcal{B}(\mathcal{P}(\widetilde \O))$-measurable. As $\mathcal{B}_c(\O ) \otimes  \mathcal{B}(\mathcal{P}(\widetilde \O)) \subset  \mathfrak{A} \left(\mathcal{B}_{c}(\O)\otimes \mathcal{B}(\mathcal{P}(\widetilde \O))\right),$  for any $c \in \mathbb{R}$, 
$$E_c:=\{(\o,q) \in \O \times \mathcal{P}(\widetilde \O): \;  \lambda (\o,q) <c\} \cap {\rm gph} \,   \Pc  \in \mathfrak{A}\left(\mathcal{B}_{c}(\O)\otimes \mathcal{B}(\mathcal{P}(\widetilde \O))\right)$$ and 
 we obtain that 
$\{\o \in \Omega:\; \Lambda(\o) <c\}={\rm proj}_{\O} E_c  \in \mathcal{B}_{c}(\O),$ using  \cite[Lemma 4.11]{BN} which relies on \cite{Lee78}. So,  $B=\{\Lambda \geq 1\} \in \mathcal{B}_{c}(\O)$. \\
Assume now that $\overline{B}$ is a $\Qc$-full measure set. 
 We  prove that $B$ is  of $\Rc$-full measure. Assume by contraposition that there exists $\tilde R \in \Rc $ such that $\tilde R (\Omega \setminus B)>0.$ 
Using \cite[Lemma 4.11]{BN}  again,  there exists $ \hat q:$ ${\rm proj}_{\O}E_1 \to  \mathfrak{P}(\widetilde \O)$ such that  $q(A,\cdot)$ is $\Bc_c(\Omega)$-measurable for all $A\in \Bc(\widetilde{\O})$ and 
$(\o,  \hat q (\cdot,\o)) \in E_1$
for all $\o \in {\rm proj}_{\O}E_1=\Omega \setminus B.$ 
As ${\rm proj}_{\O}{\rm gph} \,   \Pc=\Omega,$ the same lemma shows that there exists some 
 $\bar q \in \mathcal{S}K$ such that for all $\o \in \Omega$, 
 $\bar q (\cdot,\o) \in \Pc(\o).$ We set 
$$\tilde q (\cdot,\o):= \hat q (\cdot,\o)1_{\Omega \setminus B}(\o)+ \bar q (\cdot,\o)1_{B}(\o).$$
As $B \in \Bc_c({\O})$,  $ \tilde q  \in \mathcal{S}K $. Moreover,  as for 
$\o \in \Omega \setminus B,$  $(\o,  \hat q (\cdot,\o)) \in E_1 \subset {\rm gph} \,   \Pc,$ we conclude that 
 $\tilde q (\cdot,\o) \in \Pc(\o)$ for all $\o \in \Omega$ and that $\tilde R \otimes \tilde q  \in \Qc.$ So,
 \begin{eqnarray*}
 \tilde R \otimes \tilde q (\overline B) & = & \int_{B} \int_{\tilde \O} 1_{\overline B}(\o,\tilde \o)  \bar q (d \tilde \o,\o) \tilde R(d \o) + \int_{\Omega \setminus B}   \hat q(\overline B_{\o},\o) \tilde R(d \o) \\
  &  \leq & \tilde R(B) + \int_{\Omega \setminus B}  \l(\o, \hat q(\cdot,\o) )\tilde R(d \o) <   \tilde R(B) + \tilde R(\Omega \setminus B)=1,
 \end{eqnarray*}
as for all $\o \in \Omega \setminus B,$  $(\o,  \hat q (\cdot,\o)) \in E_1 \subset \{\lambda <1\}$ and $\tilde R (\Omega \setminus B)>0.$ This contradicts 
  the fact that $\overline B$ is  of $\Qc$-full measure and we conclude that  $B$ is  a $\Rc$-full measure set.
$\Box$\\
We derive now easily the equivalence between $\Phi \geq 0$ $ \mathcal{Q}$-q.s. and the existence of some $\Rc$-full measure set such that for all $\o$ in this  set, $\Phi(\o,\cdot) \geq 0$ $ \mathcal{Q}(\o)$-q.s. 
\begin{corollary}
\label{corlelemme}
 Assume that ${\rm gph} \,   \Pc \in \mathfrak{A} \left(\mathcal{B}_{c}(\O)\otimes \mathcal{B}(\mathcal{P}(\widetilde \O))\right)$.   
 Assume that $\Phi: \O \times \widetilde \O \to \mathbb{R}\cup \{+\infty\}$ is $\mathcal{B}_{c}(\O)\otimes \mathcal{B}(\widetilde \O)$-measurable. 
 Then there is an equivalence between:\\
 1. $\Phi \geq 0\; \Qc{\rm -q.s.}$\\
 2. There exists a $\Rc$-full measure set $\hat{\O} \in \mathcal{B}_{c}(\O)$ such that for all $\o \in \hat{\O}$
 $$\Phi(\o,\cdot) \geq 0   \; \Pc(\o){\rm -q.s.}$$
\end{corollary}
 {\sl Proof.}
To show that 1. implies 2., we apply Lemma \ref{lelemme} to  $\bar B=\{\Phi \geq 0\}$. 
The reverse implication is obtained by Fubini theorem. 
$\Box$\\

We now prove a proposition which generalizes the preceding result starting from a $\Qc_{t+1,l+1}(\o^t){\rm -q.s.}$ inequality instead of a $\Qc^{l+1}{\rm -q.s.}$ one. What is important here is that we only assume that  ${\rm gph} \,   \Qc_{l+1}$ is an analytic set    and not the analycity of   ${\rm gph}\, \Qc_{s,l+1}$  for all  $s\in \{t+2, \cdots, l+1\}.$
\begin{proposition}
\label{lelemmebis}
Fix $0 \leq t \leq  l  \leq T-1$. 
Assume that  ${\rm gph} \,   \Qc_{l+1}$ is an analytic set  on $\O^{l} \times \mathfrak{P} (\O_{l+1})$  
and  that $\Phi: \O^{l+1} \to \mathbb{R}\cup \{+\infty\}$ is a $\mathcal{B}_{c}(\O^{l} )\otimes \mathcal{B}(\O_{l+1})$-measurable function. 
Let $\widetilde{\Omega}^t \in \Bc_c(\O^t)$ be a $\Qc^t$-full measure set and 
assume that for all $\o^t \in \widetilde{\Omega}^t,$ 
$$\Phi(\o^t,\cdot) \geq 0\;\Qc_{t+1,l+1}(\o^t){\rm -q.s.}$$ 
Consider  the set-valued mapping  
$A^{t+1,l}: \O^t \twoheadrightarrow {\Omega}^{t+1,l}$ defined for all $\o^t \in \O^t$ by 
\begin{eqnarray*}
A^{t+1,l}(\o^t) & := &  \{\o^{t+1,l} \in {\Omega}^{t+1,l}: \; \Phi(\o^t,\o^{t+1,l},\cdot) \geq 0   \; \Qc_{l+1}(\o^{t},\o^{t+1,l}){\rm -q.s.}\}.
\end{eqnarray*}
Then for all  $\o^t \in \widetilde{\Omega}^t,$   $A^{t+1,l}(\o^t) \in \Bc_c(\O^{t+1,l})$ and $A^{t+1,l}(\o^t)$ is a $\Qc_{t+1,l}(\o^t)$-full measure set. 
Moreover,  ${\rm gph} \, A^{t+1,l} \in B_c(\O^{l})$ and  is a $\Qc^l$-full measure set  and 
for all $\o^l\in  {\rm gph} \, A^{t+1,l} $ 
$$\Phi(\o^l,\cdot) \geq 0\;\Qc_{l+1}(\o^l){\rm -q.s.}$$
 \end{proposition}
 {\sl Proof.}
The proof relies on two successive applications of Lemma \ref{lelemme}. 
The first one will allow to prove that $A^{t+1,l}(\o^t) \in \Bc_c(\O^{t+1,l})$ and that $A^{t+1,l}(\o^t)$ is a $\Qc_{t+1,l}(\o^t)$-full measure set for all $\o^{t} \in \widetilde{\Omega}^t.$ 
The second one will show that ${\rm gph} \, A^{t+1,l} \in B_c(\O^{l}).$ \\
{\it Step 1  $A^{t+1,l}(\o^t) \in \Bc_c(\O^{t+1,l})$  is a $\Qc_{t+1,l}(\o^t)$-full measure set for all $\o^{t} \in \widetilde{\Omega}^t.$ }\\
 Fix $\o^t \in \widetilde{\Omega}^t.$ 
We want to apply Lemma \ref{lelemme} to $\O=\O^{t+1,l}$, $\tilde \O=\O_{l+1},$ $\Rc=\Qc_{t+1,l}(\o^{t})$ and $\Pc(\o^{t+1,l})=\Qc_{l+1}(\o^{t},\o^{t+1,l}).$ 
Then, 
$q(\cdot,(\o^t,\cdot)) \in \mathcal{S}K$ if we have that $q(\cdot,(\o^t,\o^{t+1,l}))$ is a probability measure in $\mathfrak{P}(\O_{l+1})$ for all $\o^{t+1,l} \in \O^{t+1,l}$ and 
$\o^{t+1,l}\mapsto q(A,(\o^t,\o^{t+1,l}))$ is $\Bc_c(\O^{t+1,l})$-measurable for all $A\in \Bc(\O_{l+1}).$ Thus $q \in \mathcal{S}K_{l+1}(\o^t).$ 
So, 
\begin{align*}
 \mathcal{Q} & = 
\begin{multlined}[t][\textwidth-3cm]
 \{ R  \otimes q(\cdot,\o^t):\;   R \in \Qc_{t+1,l}(\o^{t}),\; q \in \mathcal{S}K_{l+1}(\o^t), \\
 q(\cdot,(\o^t,\o^{t+1,l})) \in \mathcal{Q}_{l+1}(\o^{t},\o^{t+1,l}), \;  
 \forall \; \o^{t+1,l} \in \O^{t+1,l} \} = \mathcal{Q}_{t+1,l+1}(\o^t).
 \end{multlined}
\end{align*}
We first show that ${\rm gph} \, \Pc  \in \mathfrak{A} \left(\mathcal{B}_{c}(\O^{t+1,l})\otimes \mathcal{B}(\mathcal{P}( \O_{l+1}))\right).$ 
Consider the following  $f$ $:$ $\O^{t+1,l} \times \mathfrak{P}\left(\Omega_{l+1}\right)$ $\to \Omega^{l} \times \mathfrak{P}\left(\Omega_{l+1}\right)$ be defined by 
$f(\o^{t+1,l},P):=((\o^{t},\o^{t+1,l}),P)$. Then, $f$ is Borel.  We get that 
\begin{eqnarray*}
{\rm gph} \, \Pc & = &{\rm gph} \, \Qc_{l+1}(\o^{t},\cdot) = \{(\o^{t+1,l},P),\, P \in  \Qc_{l+1}(\o^{t},\o^{t+1,l}) \} \\
& = & \{(\o^{t+1,l},P),\, ((\o^{t},\o^{t+1,l}),P) \in {\rm gph} \, \Qc_{l+1} \}=  f^{-1}({\rm gph} \, \Qc_{l+1}).
\end{eqnarray*}
So, using  \cite[Proposition 7.40 p165]{bs} ${\rm gph} \, \Pc$ is analytic. Note that we have proved that the section of an analytic set is an analytic set.\\
Moreover $
 \mathcal{A} \left(\O^{t+1,l} \otimes \mathcal{P}( \O_{l+1}) \right) \subset  
 \mathfrak{A} \left(\mathcal{B}_{c}(\O^{t+1,l})\otimes \mathcal{B}(\mathcal{P}(\O_{l+1} )\right)$ and thus we obtain that ${\rm gph} \,   \Pc  \in \mathfrak{A} \left(\mathcal{B}_{c}(\O^{t+1,l})\otimes \mathcal{B}(\mathcal{P}( \O_{l+1}))\right).$ 
 Let 
$$\overline B:=\{(\o^{t+1,l},\o_{l+1}) \in \O^{t+1,l} \times   \O_{l+1}: \;  
\Phi(\o^t,\o^{t+1,l},\o_{l+1}) \geq 0\}.$$
By assumption $\overline B$ is a $\Qc_{t+1,l+1}(\o^t)$-full measure set for all $\o^t \in  \widetilde{\O}^{t}.$ 

In order to show that $\overline B\in  \mathcal{B}_{c}(\O^{t+1,l} )\otimes \mathcal{B}(\O_{l+1})$, we prove that 
 $\Phi(\o^t,\cdot, \cdot)$ is $\mathcal{B}_{c}(\O^{t+1,l} )\otimes \mathcal{B}(\O_{l+1})$-measurable. 
 Let $h$ $:$ $\O^{t+1,l} \times \Omega_{l+1}  \to \Omega^{l+1}$ be defined by 
$h(\o^{t+1,l},\o_{l+1}):=(\o^{t},\o^{t+1,l},\o_{l+1}).$ 
Let $C \in \Bc(\mathbb{R}\cup \{+\infty\})$. Then, 
\begin{eqnarray*}
\Phi^{-1}(\o^t,\cdot, \cdot) (C)  & = & \{(\o^{t+1,l},\o_{l+1}),\, \Phi(\o^t,\o^{t+1,l},\o_{l+1}) \in C\} \\
 & = & \{(\o^{t+1,l},\o_{l+1}),\,  \Phi \circ h (\o^{t+1,l},\o_{l+1}) \in C \}= h^{-1}(\Phi^{-1}(C)).
\end{eqnarray*}
By Assumption $\Phi^{-1}(C)\in \mathcal{B}_{c}(\O^{l} )\otimes \mathcal{B}(\O_{l+1})$ and it remains to prove that 
$$h^{-1}(\mathcal{B}_{c}(\O^{l} )\otimes \mathcal{B}(\O_{l+1})) \subset \mathcal{B}_{c}(\O^{l,t+1} )\otimes \mathcal{B}(\O_{l+1}).$$
Let $A \in \mathcal{B}_{c}(\O^{l} )$ and $B \in \mathcal{B}(\O_{l+1})$. Then, $h^{-1}(A \times B)=g^{-1}(A) \times B$,   where 
$g:$ $\Omega^{t+1,l}  \to \Omega^{l}$ is defined by 
$g(\o^{t+1,l}):=(\o^{t},\o^{t+1,l})$ and  is Borel measurable. Thus, $g^{-1}(A) \times B \in \mathcal{B}_{c}(\Omega^{t+1,l} )\otimes \mathcal{B}(\O_{l+1})$  
and we have proved that 
$$h^{-1}(\mathcal{B}_{c}(\O^{l} )\times \mathcal{B}(\O_{l+1})) \subset \mathcal{B}_{c}(\O^{t+1,l} )\otimes \mathcal{B}(\O_{l+1}).$$
The transport lemma shows that $$h^{-1}(  \mathcal{B}_{c}(\O^{l} )\otimes \mathcal{B}(\O_{l+1}))=\sigma(h^{-1}( \mathcal{B}_{c}(\O^{l} )\times \mathcal{B}(\O_{l+1})))
\subset \mathcal{B}_{c}(\Omega^{t+1,l} )\otimes \mathcal{B}(\O_{l+1}).$$
Thus, we have proved that $\overline B\in  \mathcal{B}_{c}(\O^{t+1,l} )\otimes \mathcal{B}(\O_{l+1})$ and we can apply Lemma \ref{lelemme}. The set $B$ of Lemma \ref{lelemme} (see \eqref{BB}) is  given by 
\begin{eqnarray*}
B& := & \left\{\o^{t+1,l} \in \Omega^{t+1,l}: \;  q\left(\{\o_{l+1}\in \O_{l+1}: \;  \Phi(\o^t,\o^{t+1,l},\o_{l+1}) \geq 0\right)=1, \right.\\
& & \left. \forall \; q \in \Qc_{l+1}(\o^{t},\o^{t+1,l})\right\}=A^{t+1,l}(\o^t) \in \Bc_c(\O^{t+1,l}).
\end{eqnarray*}
As for all $\o^t \in \widetilde{\Omega}^t,$ $\overline B$ is of $\Qc_{t+1,l+1}(\o^t)$-full measure, $B=A^{t+1,l}(\o^t)$ is a $\Qc_{t+1,l}(\o^t)$-full measure set.  \\

{\it Step 2 ${\rm gph} \, A^{t+1,l} \in B_c(\O^{l}).$}\\  
We apply Lemma \ref{lelemme} with $\O=\O^{l}$, $\tilde \O=\O_{l+1},$ $\Rc=\Qc^{l},$ $\Pc(\o^{l})=\Qc_{l+1}(\o^{l})$ and 
$$\overline B:=\{(\o^{l},\o_{l+1}) \in \O^{l} \times   \O_{l+1}: \;  
\Phi(\o^{l},\o_{l+1}) \geq 0\}. $$
By assumption $\overline B \in  \mathcal{B}_{c}(\O^{l} )\otimes \mathcal{B}(\O_{l+1})$ and ${\rm gph} \,   \Pc$ is an analytic set, so we can apply 
Lemma \ref{lelemme} and recalling  \eqref{BB}, we get that 
\begin{eqnarray}
 \nonumber
B & := & \left\{\o^{l} \in \Omega^{l}: \;  q\left(\{\o_{l+1}\in \O_{l+1}: \;  \Phi(\o^{l},\o_{l+1}) \geq 0\right)=1, 
\forall \; q \in \Qc_{l+1}(\o^{l})\right\} \\
\label{arlm}
& =   & 
 \{\o^{l}\in \Omega^{l}:\, \Phi(\o^{l},\cdot) \geq 0  \,\Qc_{l+1}(\o^{l}){\rm -q.s.}\}\\ 
 \nonumber
 &= & \{(\o^{t},\o^{t+1,l})\in \Omega^{l}:\; \o^{t+1,l}\in A^{t+1,l}(\o^t)\}= {\rm gph} \, A^{t+1,l} \in \mathcal{B}_{c}(\O^{l}). 
\end{eqnarray} 

{\it Step 3}  $ {\rm gph} \, A^{t+1,l}$ is a $\Qc^{l}$-full measure set.\\
Let  $P^* =P\otimes q\in \Qc^{l}$ where 
$P \in \Qc^{t}$ and $q(\cdot ,\o^{t})\in \Qc_{t+1,l}(\o^{t})$ for all $\o^{t} \in {\O}^{t}.$ 
\begin{eqnarray}
\nonumber
 P^* ({\rm gph} \,  A^{t+1,l}) & = &  \int_{\O^{l}}  1_{{\rm gph} \,  A^{t+1,l}}(\o^{l})P^*(d \o^{l})\\
  & = &
  \nonumber
    \int_{\O^{t}} \int_{\O^{t+1,l}} 1_{  A^{t+1,l}(\o^{t})}(\o^{t+1,l})q(d  \o^{t+1,l},\o^{t})P(d \o^{t}) \\
\label{life}
 & = &    \int_{\widetilde{\O}^{t}}  q( A^{t+1,l}(\o^{t}),\o^{t})P(d \o^{t}) =1.
 \end{eqnarray}
The first equality in \eqref{life} holds true because 
 $\widetilde{\O}^{t}$ is a $\Qc^{t}$-full  measure set and $P \in \Qc^{t}$.  
 The second equality in \eqref{life} follows from step 1:   for all $\o^{t} \in\widetilde{\O}^{t},$ $ A^{t+1,l}(\o^{t})$ is of $\Qc_{t+1,l}(\o^{t})$-full measure and $q(\cdot ,\o^{t}) \in \Qc_{t+1,l}(\o^{t})$. The proof is finished recalling \eqref{arlm}. $\Box$
\begin{acknowledgements}
The author would like to thank Romain Blanchard for helpful discussions.
\end{acknowledgements}
%
\section*{Conflict of interest}
The author declares that she has no conflict of interest.
\bibliographystyle{spmpsci}      
\bibliography{biblioRomain}   

\end{document}